\documentclass[reprint,nofootinbib,aps,superscriptaddress,amsmath,amssymb,onecolumn,11pt]{revtex4-2}

\usepackage{graphicx}
\usepackage{dcolumn}
\usepackage{bm}
\usepackage{braket}
\usepackage{breqn}
\usepackage{mathrsfs}
\usepackage{color}

\usepackage[colorlinks=true,linkcolor=red,citecolor=blue]{hyperref}

\begin{document}

\title{Energy Extraction via Magnetic Reconnection in Konoplya-Rezzolla-Zhidenko Parametrized Black Holes}

\author{Shao-Jun Zhang}
\email{sjzhang@zjut.edu.cn}
\affiliation{Institute for Theoretical Physics and Cosmology$,$ Zhejiang University of Technology$,$ Hangzhou 310032$,$ China}
\affiliation{United Center for Gravitational Wave Physics$,$ Zhejiang University of Technology$,$ Hangzhou 310032$,$ China}
\date{\today}

\begin{abstract}
Recently, Comisso and Asenjo proposed a novel mechanism for harnessing energy from black holes through magnetic reconnection. Our study focuses on exploring the utilization of this mechanism on Konoplya-Rezzolla-Zhidenko (KRZ) parametrized black holes to assess the impact of deformation parameters on energy extraction. Among the various parameters, $\{\delta_1, \delta_2\}$ are identified as the most important factors affecting the physics under consideration. The influence of these two parameters on the ergoregion, circular geodesics in the equatorial plane, and energy extraction from the KRZ black holes through magnetic reconnection is carefully analyzed. Results indicate that deviations from the Kerr metric have a notable influence on the energy extraction process. Particularly, energy extraction is enhanced with more negative ${\delta_1}$ or more positive ${\delta_2}$ within the range constrained by current astronomical observations, resulting in significantly higher maximum power and efficiency compared to the Kerr model. Moreover, when ${\delta_2}$ is sufficiently negative, extracting energy through this mechanism becomes increasingly challenging, necessitating an exceptionally high black hole spin.
\end{abstract}


\maketitle
\section{Introduction}

Black holes, unique and compact objects predicted by Einstein's theory of gravity, have a significant impact on contemporary physics. Recent advances in astronomical observations have ushered in a golden age for delving into the physics of black holes \cite{LIGOScientific:2016aoc, LIGOScientific:2016sjg, LIGOScientific:2017bnn, LIGOScientific:2017ycc, LIGOScientific:2020iuh, EventHorizonTelescope:2019dse}. These mysterious objects are thought to be responsible for several energetic astrophysical phenomena, such as gamma-ray bursts (GRBs) and relativistic jets in active galactic nuclei (AGNs). The energy discharged during these phenomena may originate from the gravitational potential energy liberated as matter plunges into the black hole, or from the inherent energy of the black hole itself.

According to general relativity (GR), rotating black holes have significant amounts of energy that can be harnessed. The second law of black hole thermodynamics states that the maximum energy that can be extracted from a rotating black hole is about $0.29 Mc^2$ \cite{Christodoulou:1970wf}, where $M$ is the mass of the black hole and $c$ is the speed of light in vacuum. This energy originates from the black hole's rotational energy, prompting an intriguing inquiry into the processes involved in extracting such a significant fraction of the black hole's energy. 

The first plausible mechanism for energy extraction from black holes was proposed by Penrose in 1969, now known as the Penrose process \cite{Penrose:1969pc}. The Penrose process involves the splitting of a particle in the ergoregion into two particles. Measured from infinity, one of these particles may carry negative energy, while the other carries excess positive energy. The extraction of energy from a black hole occurs when the negative energy particle is absorbed by the black hole and the positive energy particle escapes. However, the Penrose process is considered impractical because it requires the two newborn particles to have relative velocities greater than 1/2 the speed of light, an event thought to be rare in real astrophysical processes \cite{Bardeen:1972fi, Wald:1974kya}. Inspired by Penrose's work, physicists have proposed several alternative mechanisms for harnessing energy from black holes, including superradiant scattering \cite{Teukolsky:1974yv}, the collisional Penrose process \cite{piran1975high}, the Blandford-Znajek (BZ) process \cite{Blandford:1977ds}, and the magnetohydrodynamic (MHD) Penrose process \cite{1990ApJ...363..206T}. Among these mechanisms, the BZ process is currently regarded as the most promising for elucidating GRBs \cite{Lee:1999se, Tchekhovskoy:2008gq, Komissarov:2009dn} and relativistic jets in AGNs \cite{McKinney:2004ka, Hawley:2005xs, Komissarov:2007rc, Tchekhovskoy:2011zx}.

Recently, based on an exploratory study \cite{Koide:2008xr}, a novel mechanism for black hole energy extraction has been proposed by Comisso and Asenjo \cite{Comisso:2020ykg}. In this mechanism, energy extraction is realized by a process of magnetic reconnection in the ergoregion of a rotating black hole. The process begins with the magnetic field around the black hole forming antiparallel configurations near the equatorial plane due to the black hole's rotation \cite{Parfrey:2018dnc, Komissarov:2005wj, East:2018ayf, Ripperda:2020bpz}. The jumps in the magnetic field direction induce the formation of current sheets between them, which are destroyed by plasmoid instability when their aspect ratio exceeds a critical value \cite{comisso2016General, Uzdensky:2010ts, Comisso:2017arh}, resulting in the formation of plasmoids/flux ropes that drive fast magnetic reconnection to convert magnetic energy into plasma kinetic energy \cite{daughton2009Transition,bhattacharjee2009Fast}. Eventually, the plasma is expelled from the reconnection layer and the tension in the magnetic field is released as the plasma flows out. The magnetic field lines are then reextended by the dragging of the black hole to form an antiparallel configuration, and the current sheet is induced again to repeat the magnetic reconnection process. During each magnetic reconnection, the plasma in the current sheet splits into two parts, one corotating with the black hole and the other counter-rotating. The corotating part is accelerated while the counter-rotating part is decelerated. Similar to the Penrose process, if the decelerated part has negative energy and is absorbed by the black hole, while the accelerated part escapes, the accelerated part ends up carrying extra energy, thus realizing energy extraction from the black hole. This extra energy comes from the black hole's rotational energy. 

Astronomical observations have confirmed that there are often different scales of magnetic fields around black holes due to the presence of accretion matter or companion stars. For example, the supermassive black hole Sagittarius A* is accompanied by the magnetar SGR J1745-2900 \cite{Mori:2013yda, Kennea:2013dfa, Eatough:2013nva, Olausen:2013bpa}, and a strong magnetic field is also present near the event horizon of the black hole in M87* \cite{EventHorizonTelescope:2021srq}. Therefore, the process of magnetic reconnection of black holes can be expected to occur routinely. In fact, in the BZ mechanism, energy extraction is also realized with the help of the magnetic field around the black hole. The Comisso-Asenjo mechanism is distinguished from the BZ mechanism by its bursty nature and reliance on nonzero particle inertia. Studies by Comisso and Asenjo suggest that this mechanism may be more efficient than the BZ process in certain parameter spaces, making it a promising candidate for energy extraction from black holes. 

On the other hand, to test GR and the Kerr hypothesis, it is necessary to construct other non-Kerr rotating black holes. There are usually two ways to construct them, the top-down approach and the bottom-up approach. In the top-down method, one starts from a modified gravitational theory and obtains exact rotating black hole solutions by solving the gravitational field equations. However, solving modified gravitational field equations is generally very challenging due to their complexity. Only in a few cases or under certain approximations (e.g., slow rotation) can exact analytical or numerical rotating black hole solutions be obtained, e.g., the Einstein-dilaton-Gauss-Bonnet black holes \cite{Kanti:1995vq, Ayzenberg:2014aka, Maselli:2015tta, Kleihaus:2015aje, Kokkotas:2017ymc}, the Chern-Simons black holes \cite{Yunes:2009hc, Yagi:2012ya, McNees:2015srl, Delsate:2018ome}, and the Kerr-Sen black holes \cite{Sen:1992ua}. Conversely, the bottom-up method involves constructing parametrized black holes by generalizing the Kerr metric based on considerations like symmetries and regularity, which are independent of any specific gravity theory \cite{Vigeland:2011ji, Johannsen:2013szh, Konoplya:2016jvv, Papadopoulos:2018nvd, Carson:2020dez}. These non-Kerr rotating black holes usually contain several deformation parameters introduced to characterize the deviations from the Kerr black hole. Several of them align well with present-day astronomical observations, indicating that their presence cannot be discounted. Consequently, they may offer viable alternatives to the Kerr metric given the current level of accuracy in astronomical observations. Studies of black hole energy extraction via the Comisso-Asenjo mechanism have also been extended to these non-Kerr rotating black holes, such as spinning braneworld black holes \cite{Wei:2022jbi}, the Johannsen-Psaltis metric \cite{Liu:2022qnr}, Kerr-Sen and Kiselev black holes \cite{Carleo:2022qlv},  rotating black holes with broken Lorentz symmetry \cite{Khodadi:2022dff}, spinning regular black hole \cite{Li:2023nmy}, spinning hairy black hole \cite{Li:2023htz}, Kerr-MOG black holes \cite{Khodadi:2023juk}, Kerr-Newman-MOG black holes \cite{Shaymatov:2023dtt} and Kerr-de Sitter black holes \cite{Wang:2022qmg}. These studies have demonstrated that deviations from the Kerr black hole significantly impact the energy extraction process.

Inspired by these studies, our main goal in this paper is to study the energy extraction process via the Comisso-Asenjo mechanism for KRZ parametrized black holes \cite{Konoplya:2016jvv}. The KRZ metric is proposed to describe generic stationary and axisymmetric black holes. It has several subtle advantages over other parametrized metrics.  In the weak-gravity regime, it can be well-aligned with Kerr black holes and thus passes the current astronomical observational constraints, while in the near-horizon strong-gravity regime, it needs only a few dominant parameters to sufficiently describe generic deviations from the Kerr black hole. Moreover, the KRZ metric can be mapped to many top-down metrics when the deformation parameters take on some specific values. Due to these merits, it has been utilized to test GR and the Kerr hypothesis using current astronomical data, leading to the examination of constraints on the deformation parameters \cite{Ni:2016uik, Nampalliwar:2019iti, Li:2023qcu, Abdikamalov:2021zwv, Konoplya:2021qll, Shaymatov:2023jfa}. This study primarily focuses on the impact of these deformation parameters, which signify deformations near the horizon, on the magnetic reconnection process and subsequent energy extraction. Given that magnetic reconnection occurs in the near-horizon ergoregion, it is anticipated that these deformation parameters will exert notable influences on this process. Future precise astronomical observations, particularly in the strong field region near the horizon, may facilitate the testing of Kerr's hypothesis by imposing constraints on these deformation parameters through this process.

The paper is organized as follows. In Sec. II, a brief overview of the KRZ-parametrized black hole is provided, along with an examination of how the deformation parameters impact the ergoregion. Section III delves into a thorough analysis of the circular geodesic motion of the plasma in the equatorial plane. Section IV investigates the extraction of energy from the black hole via the Comisso-Asenjo mechanism. The last section is the summary and conclusions. Throughout this work, we utilize the units $c = G =1$, where $c$ and $G$ are the speed of light in vacuum and the Newton gravitational constant, respectively.

\section{The KRZ parametrized black hole}

In Boyer-Lindquist coordinates, the KRZ metric reads \cite{Konoplya:2016jvv,Ni:2016uik,Nampalliwar:2019iti},
\begin{align}
	ds^2 = - \frac{N^2 - W^2 \sin^2 \theta}{K^2} dt^2 - 2 W r \sin^2 \theta dt d \phi + K^2 r^2 \sin^2 \theta d \phi^2 + \frac{\Sigma B^2}{N^2} dr^2 + \Sigma r^2 d \theta^2,  
\end{align} 
where
\begin{align}
	& N^2 = \left(1 - \frac{r_0}{r}\right) \left(1- \frac{\epsilon_0 r_0 }{r} + (k_{00} - \epsilon_0 ) \frac{r_0^2}{r^2} + \frac{\delta_1 r_0^3}{r^3}\right) + \left(\frac{a_{20} r_0^3}{r^3} +  \frac{a_{21} r_0^4}{r^4} + \frac{k_{21} r_0^3}{r^3 \left(1 + \frac{k_{22} \left(1- \frac{r_0}{r}\right)}{1 + k_{23} \left(1 - \frac{r_0}{r}\right)}\right)}\right) \cos^2 \theta,\\
	& B = 1 + \frac{\delta_4 r_0^2 }{r^2} + \frac{\delta_5 r_0^2 }{r^2} \cos^2 \theta,\\
    & W = \frac{1}{\Sigma} \left(\frac{\omega_{00} r_0^2}{r^2} + \frac{\delta_2 r_0^3 }{r^3} + \frac{\delta_3 r_0^3 }{r^3} \cos^2 \theta \right),\\
    & K^2 = 1 + \frac{a W}{r} + \frac{1}{\Sigma } \left(\frac{k_{00} r_0^2}{r^2} + \left(\frac{k_{20} r_0^2}{r^2} + \frac{k_{21} r_0^3}{r^3 \left(1 + \frac{k_{22} \left(1- \frac{r_0}{r}\right)}{1 + k_{23} \left(1 - \frac{r_0}{r}\right)}\right)} \right) \cos^2 \theta \right),\\
	& \Sigma = 1 + \frac{a^2}{r^2} \cos^2 \theta,
\end{align}
where $a = J / M^2$ is the dimensionless spin parameter,  $r_0 = 1 + \sqrt{1 - a^2} $ is the radius of the event horizon in the equatorial plane and 
\begin{align}
	&\epsilon_0 = \frac{2 - r_0}{r_0}, \quad a_{20} = \frac{2 a^2}{r_0^3},\quad a_{21} = - \frac{a^4}{r_0^4} + \delta_6, \quad \omega_{00} = \frac{2 a}{r_0^2},\nonumber\\
	&k_{00} = k_{23} = \frac{a^2}{r_0^2},\quad k_{20} = 0,\quad k_{21} = \frac{a^4}{r_0^4} - \frac{2 a^2}{r_0^3} - \delta_6,\quad k_{22} = - \frac{a^2}{r_0^2}.  
\end{align}
Note that in the metric, the black hole mass has been set to be $M=1$, so all physical quantities are measured in units of $M$ in the following. Here $\{\delta_i\} (i=1,2,\cdots,6)$ are six deformation parameters introduced to characterize deviations away from the Kerr metric: $\delta_1$ corresponds to deformations of $g_{tt}$, $\delta_2$ and $\delta_3$ correspond to rotational deformations of the metric, $\delta_4$ and $\delta_5$ correspond to deformations of $g_{rr}$, and $\delta_6$ corresponds to deformations of the event horizon. In the limit of all $\delta_i \rightarrow 0$, the KRZ metric reproduces the Kerr metric exactly. Note that in the metric the sign of $a$ can be absorbed into $\{\delta_2, \delta_3\}$ and by redefining $t$, so we will only consider $a>0$. Moreover, as we see later, among the six deformation parameters, $\{\delta _1, \delta _2\}$ are the most important for the physics we are going to study. So, for simplicity we will set $\delta _3 = \delta _4 = \delta _5 = \delta _6 =0$ in the following and only consider the influences of $\{\delta _1, \delta _2\}$. The constraint on $\{\delta _1, \delta _2\}$ from x-ray observations of the supermassive BH in Ark 564 is \cite{Nampalliwar:2019iti}
\begin{align}
 -0.27 < \delta_1 < 0.28, \quad -0.37 < \delta_2 < 0.22. \label{ParameterRegion}
\end{align}

\begin{figure}[!htbp]
	\includegraphics[width=0.45\textwidth]{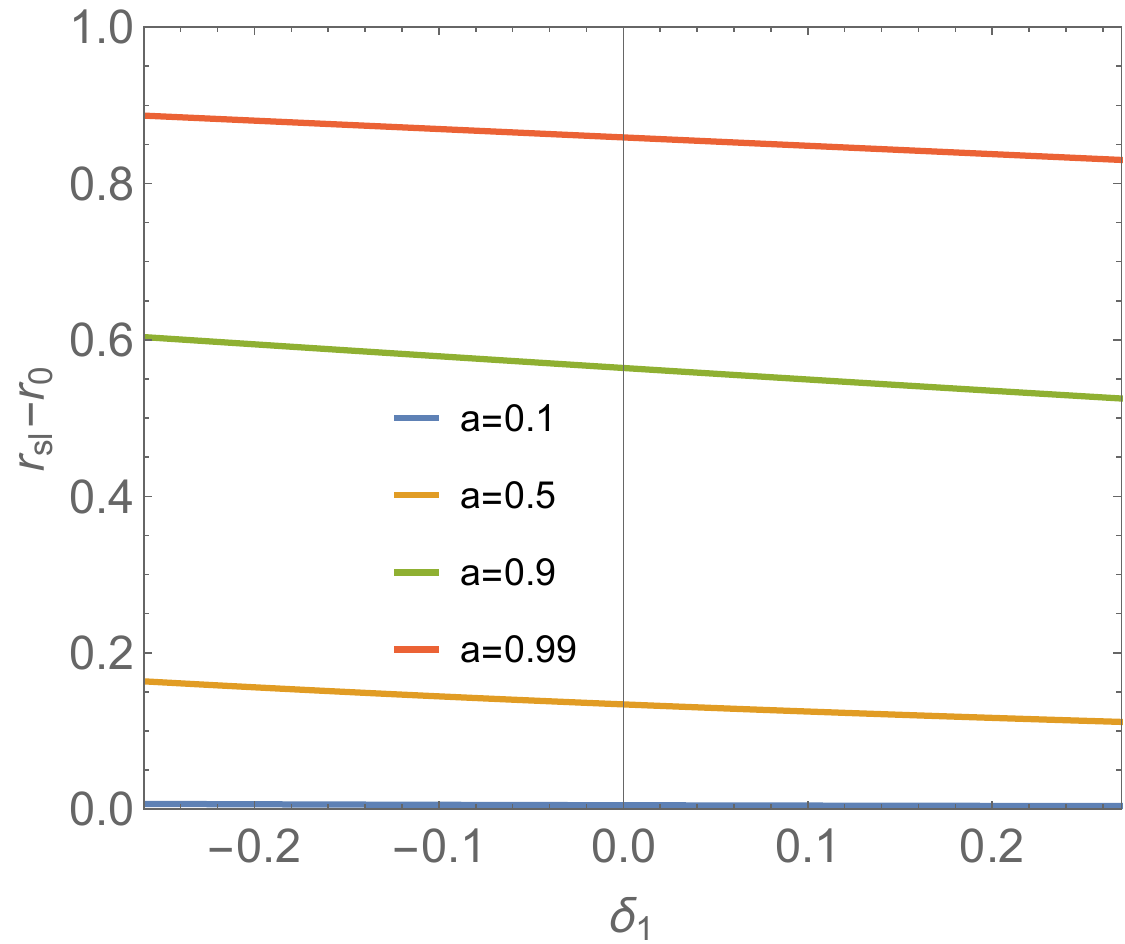}\quad
	\includegraphics[width=0.45\textwidth]{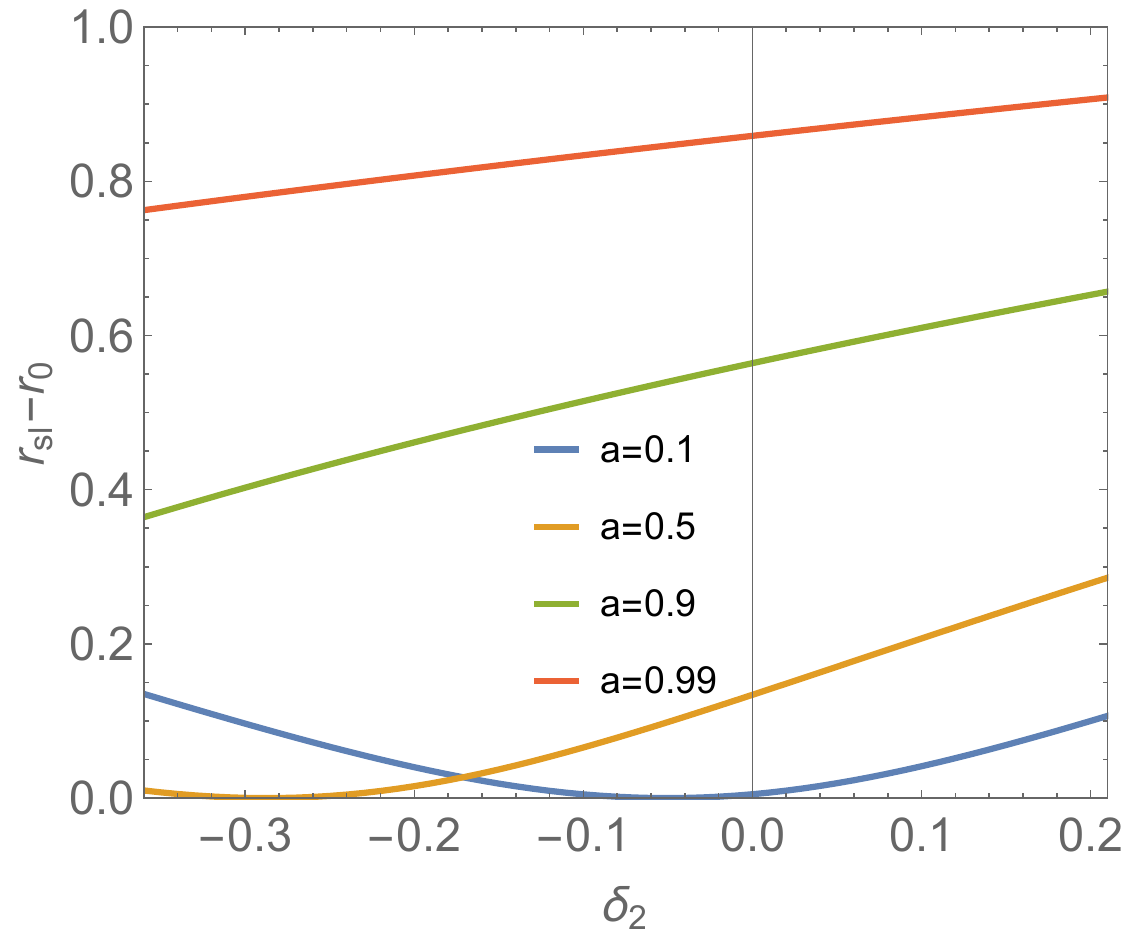}
	\caption{Ergosphere in the equatorial plane $r=r_{sl}$ as a function of $\{\delta_1, \delta_2\}$ for various $a$. The event horizon in the equatorial plane is located at $r=r_0=1+\sqrt{1-a^2}$. In the left panel $\delta_2=0$, while in the right panel $\delta_1=0$.} \label{ErgoregionEquatorial}
\end{figure}

The location of the event horizon is given by the root of the equation
\cite{Konoplya:2016jvv}
\begin{align}
	N^2 (r, \theta) =0.\label{EventHorizon}
\end{align}
For nonvanishing deformation parameters, the above equation determines the radius of the event horizon as a function of the polar angle $\theta$. The ergoregion, where magnetic reconnection occurs, lies between the event horizon and the ergosphere (the static limit surface) and is defined as
\begin{align}
	0< N^2 (r, \theta) < W^2 \sin^2 \theta.\label{Ergoregion}
\end{align}
In the equatorial plane $\theta = \pi /2$, the event horizon is at $r=r_0$, while the location of the ergosphere $r=r_{\rm sl}$ is only affected by $\{\delta_1, \delta_2\}$. Figure \ref{ErgoregionEquatorial} shows the ergosphere $r=r_{\rm sl}$ in the equatorial plane as a function of $\{\delta_1, \delta_2\}$ for various values of the spin $a$. It can be seen that the deformation parameters $\{\delta_1, \delta_2\}$ have a significant influence on the shape of the ergoregion. From the left panel, it can be seen that a larger $a$ results in a larger ergoregion. And for a fixed $a$, $r_{\rm sl}$ is a monotonically decreasing function of $\delta_1$, meaning that negative $\delta_1$ enlarges the ergoregion while positive $\delta_1$ shrinks it. The effect of $\delta_2$ on $r_{\rm sl}$ is more complicated, as shown in the right panel. For fixed and large $a$, $r_{\rm sl}$ is a monotonically increasing function of $\delta_2$, meaning that positive $\delta_2$ will expand the ergoregion, while negative $\delta_2$ will contract it. While for fixed and small $a$, $r_{\rm sl}$ is a convex function of $\delta_2$, meaning that large negative $\delta_2$ will also expand the ergoregion in this case. Note that if $\delta_2$ is sufficiently negative, a higher (but still small) spin $a$ is not necessary to produce a larger ergoregion (e.g., the ergoregion for $a=0.5$ is smaller than that for $a=0.1$ when $\delta_2$ is large negative).

\section{Circular geodesics motion in equatorial plane}

Magnetic reconnection is related to the motion of charged plasma around the compact object. Under the "force-free" assumption, the electromagnetic interaction between plasma vanishes and the particles follow geodesics which can be determined by the Lagrangian
\begin{equation}\label{Lagrangian}
	{\cal L} = \frac{1}{2} g_{\mu \nu} \dot{x}^\mu \dot{x}^\nu,
\end{equation}
where dot means the derivative with respect to some affine parameter. From the conservation of the rest mass, ${\cal L}= -1 /2$, we have \begin{equation}
	g_{rr} \dot{r}^2 + g_{\theta \theta} \dot{\theta}^2 = V_{\rm eff} (r, \theta),
\end{equation}
where the effective potential is
\begin{equation}\label{EffectivePotential-1}
	V_{\rm eff} (r, \theta)= - \left(1 + g_{tt} \dot{t}^2 + 2 g_{t \phi} \dot{t} \dot{\phi} + g_{\phi \phi} \dot{\phi}^2\right).
\end{equation}

Note that $L$ does not depend on either $t$ or $\phi$ explicitly, so both the specific energy $E$ and the $z$-component angular momentum $L_z$ of the particles are conserved along geodesics,
\begin{align}
	E &\equiv - \frac{\partial L}{\partial \dot{t}} = - g_{tt} \dot{t} - g_{t \phi } \dot{\phi},\\
	L_z & \equiv \frac{\partial L }{\partial \dot{\phi}} = g_{\phi \phi } \dot{\phi } + g_{t \phi } \dot{t}.
\end{align}
From the above equations, we then have
\begin{align}
	\dot{t} &= \frac{g_{\phi \phi } E + g_{t \phi } L_z}{g_{t \phi }^2-g_{tt} g_{\phi \phi }} , \label{tEq}\\
	\dot{\phi} &= - \frac{g_{t \phi } E + g_{t t } L_z}{g_{t \phi }^2 -g_{tt} g_{\phi \phi } } . \label{phiEq}
\end{align}  
With these relations, the effective potential (\ref{EffectivePotential-1}) can be rewritten as
\begin{equation}\label{EffectivePotential-2}
	V_{\rm eff} (r, \theta)= \frac{g_{\phi \phi} E^2 + 2 g_{t \phi} E L_z + g_{tt} L_z^2}{g_{t \phi }^2-g_{tt} g_{\phi \phi }} - 1.
\end{equation}

For circular orbits in the equatorial plane $\theta = \pi / 2$, the effective potential should satisfy the following three conditions:
\begin{align}
	V_{\rm eff} =0,\quad \partial_r V_{\rm eff} =0,\quad \partial_\theta V_{\rm eff} =0.
\end{align}
The last condition can always be satisfied due to the reflection symmetry of the spacetime with respect to the equatorial plane. With the first two conditions, one can obtain the Keplerian angular velocity $\Omega_K \equiv \dot{\phi} / \dot{t}$, the specific energy $E$ and the $z$-component angular momentum $L_z$ of the particle,
\begin{align}\label{OrbitQuantities}
	\Omega_K &= \frac{-\partial_r g_{t \phi} \pm \sqrt{(\partial_r g_{t \phi} )^2 - (\partial_r g_{tt}) (\partial_r g_{\phi \phi}) } }{\partial_r g_{\phi \phi } },\\
	E &= -\frac{g_{tt} + g_{t \phi} \Omega _K}{\sqrt{- g_{tt} - 2 g_{t \phi} \Omega _K - g_{\phi \phi} \Omega_K^2}},\\
	L_z &= \frac{g_{t \phi} + g_{\phi \phi} \Omega _K}{\sqrt{- g_{tt} - 2 g_{t \phi} \Omega _K - g_{\phi \phi} \Omega _K^2}},
\end{align}
where the sign $+$ stands for corotating orbits and $-$ for counter-rotating orbits. Note that of the six deformation parameters, only $\{\delta_1, \delta_2\}$ enter the three equations above. Given the values of the spin $a$ and the deformation parameters $\{\delta_1, \delta_2\}$, the above equations determine $\Omega_K, E$ and $L_z$ as a function of the radius of the circular orbit. The circular orbit is stable under perturbations if $\partial_r^2 V_{\rm eff} \leq 0$ and $\partial_\theta^2 V_{\rm eff} \leq 0$. A radially stable circular orbit exists from infinity to the innermost (radially) stable circular orbit (ISCO), which is given by the condition
\begin{align}\label{ISCO}
	\partial_r^2 V_{\rm eff} = 0. 
\end{align}
With the above condition, the radius of ISCO can be derived for given values of the spin $a$ and $\{\delta_1, \delta_2\}$. 

\begin{figure}[!htbp]
	\includegraphics[width=0.45\textwidth]{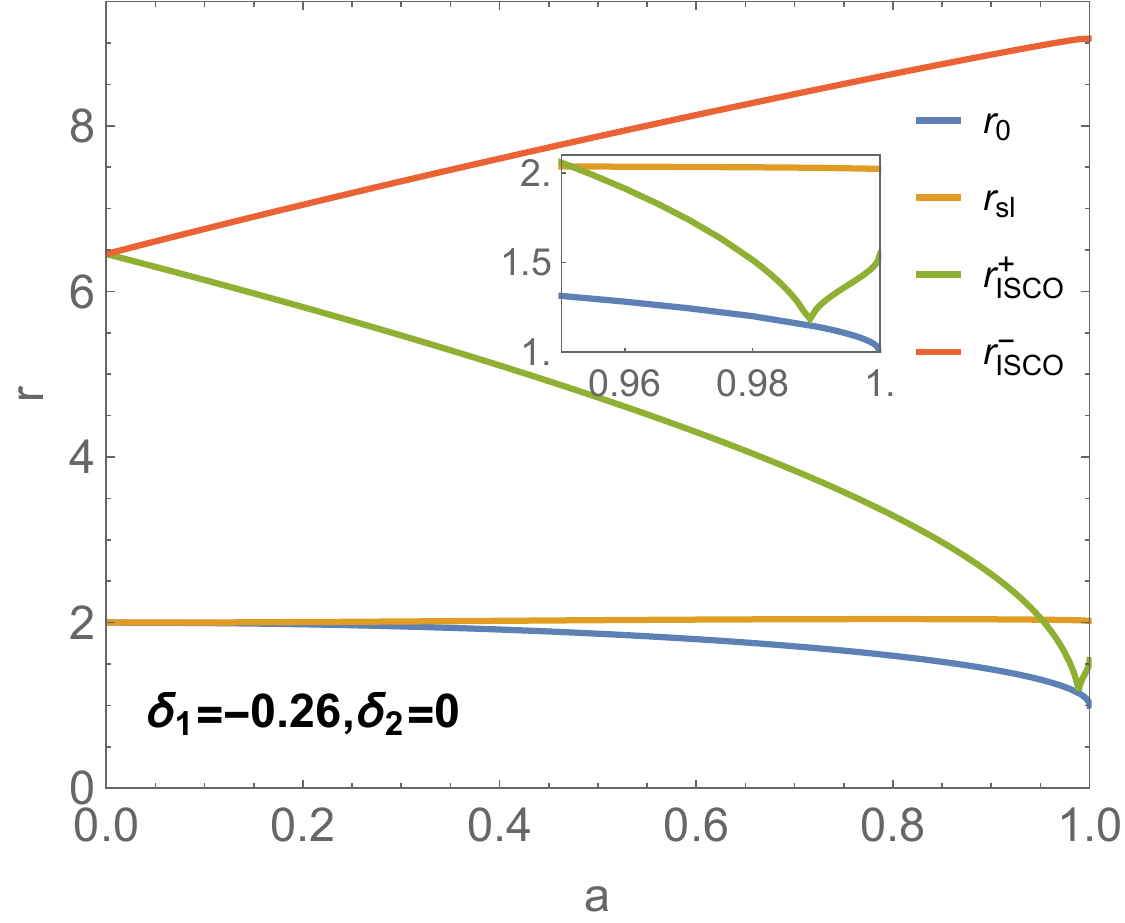}\quad
	\includegraphics[width=0.45\textwidth]{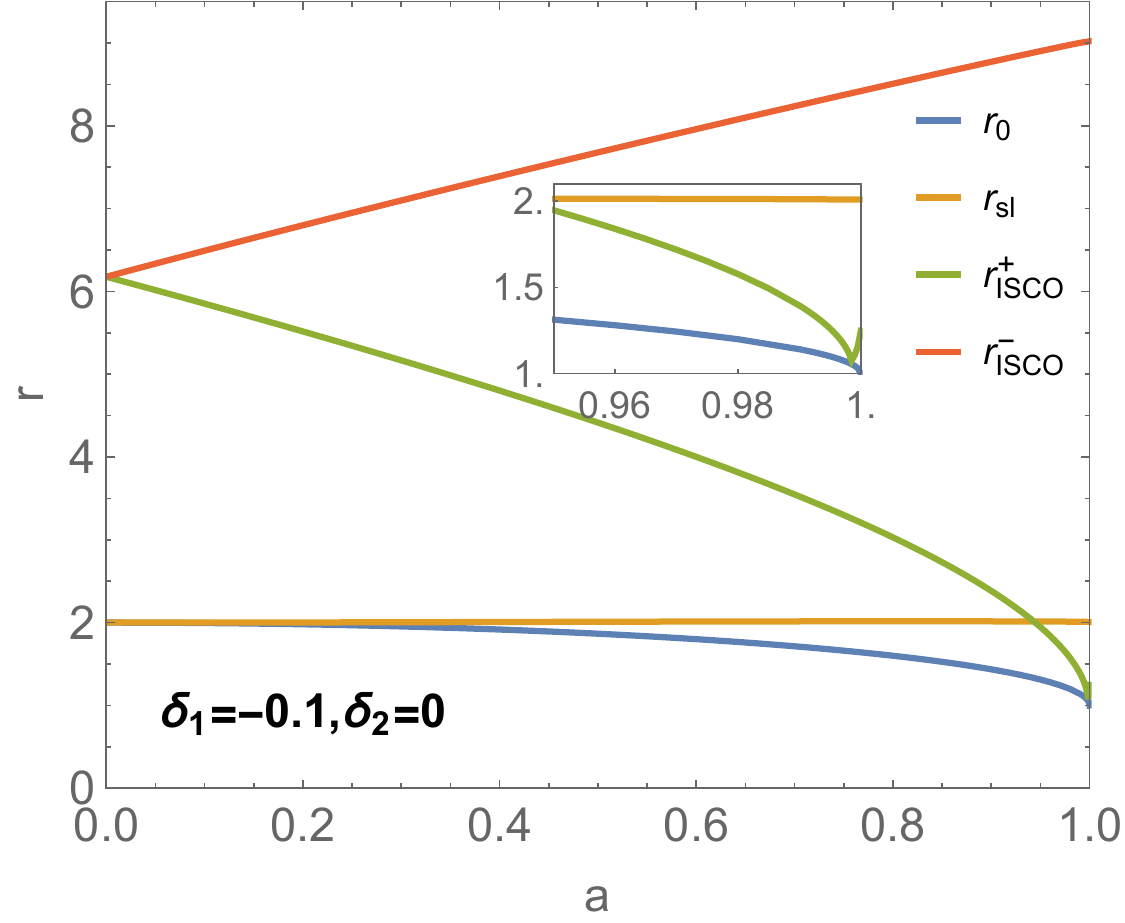}\\
	\vspace{0.5cm}
	\includegraphics[width=0.45\textwidth]{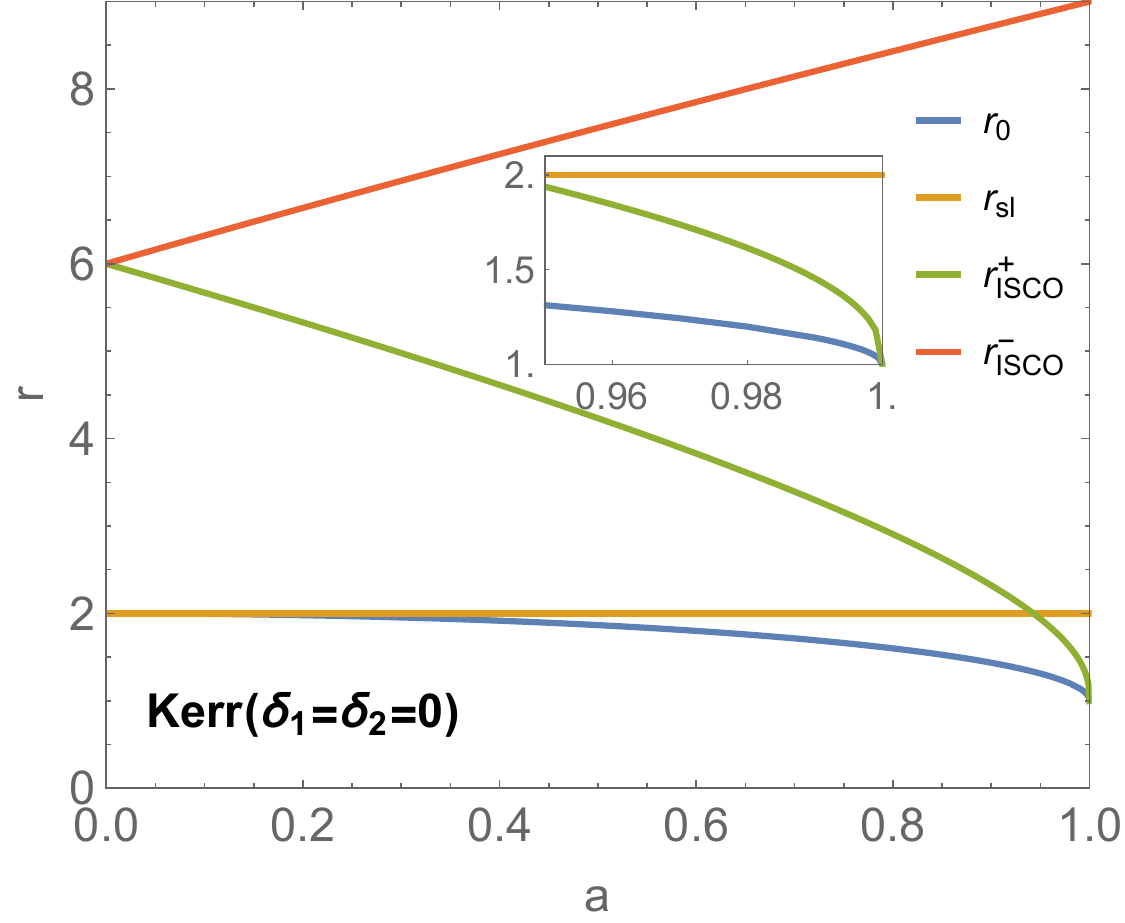}\quad
	\includegraphics[width=0.45\textwidth]{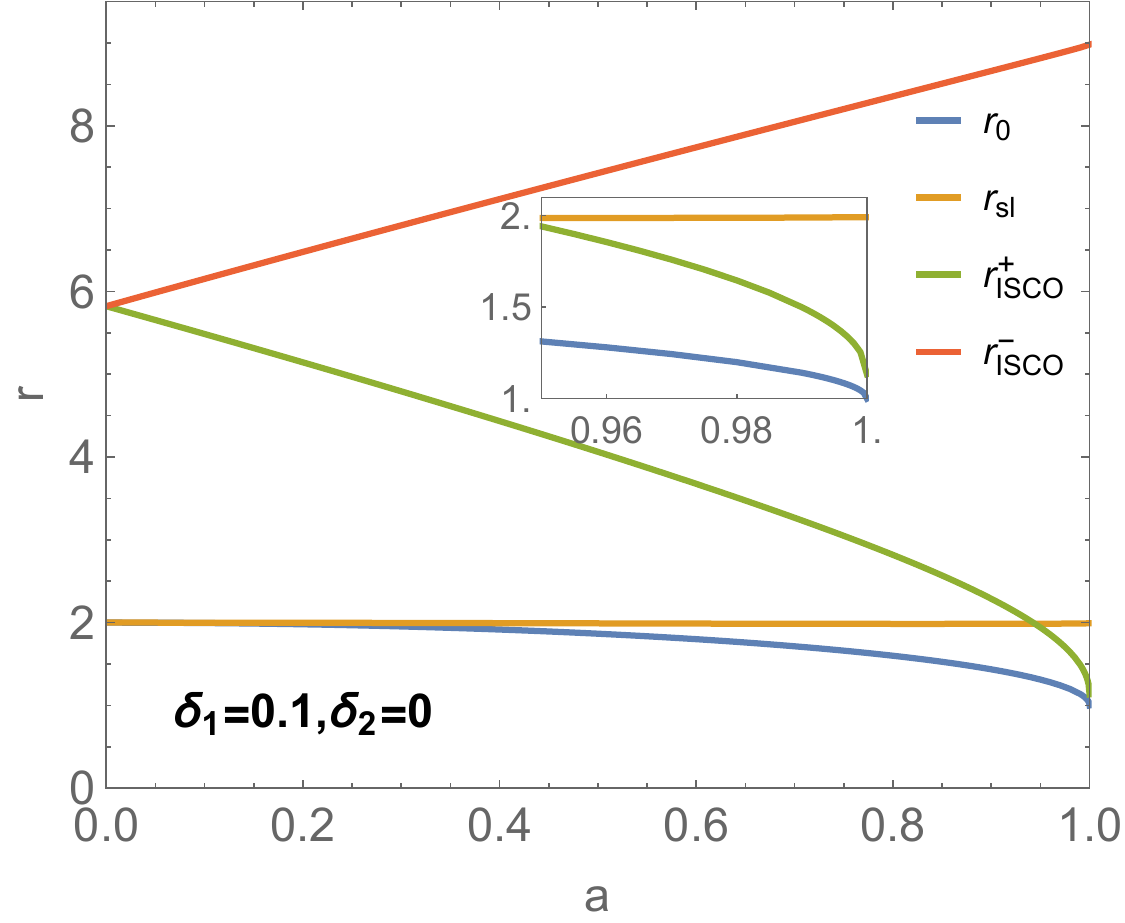}\\
	\vspace{0.5cm}
	\includegraphics[width=0.45\textwidth]{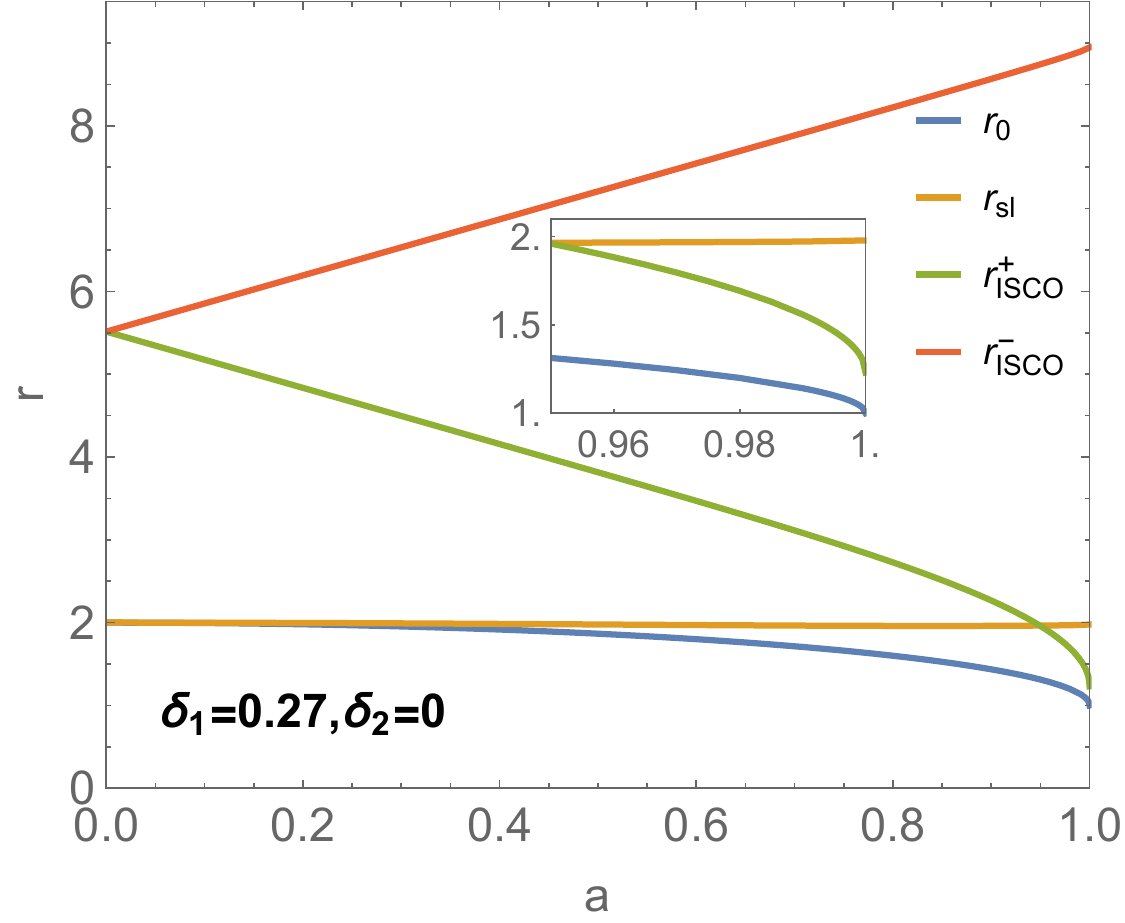}
	\caption{Radius of innermost stable circular orbit as a function of the spin $a$. Here, we fix $\delta_2 =0$ and study the influence of $\delta_1$. The Kerr case $(\delta_1=\delta_2=0)$ is also shown for comparison. $r^+_{\rm ISCO}$ and $r^-_{\rm ISCO}$ represent the corotating and counter-rotating orbits, respectively. $r_0$ is the event horizon and $r_{sl}$ is the ergosphere.} \label{ISCODelta1}
\end{figure}

\begin{figure}[!htbp]
	\includegraphics[width=0.45\textwidth]{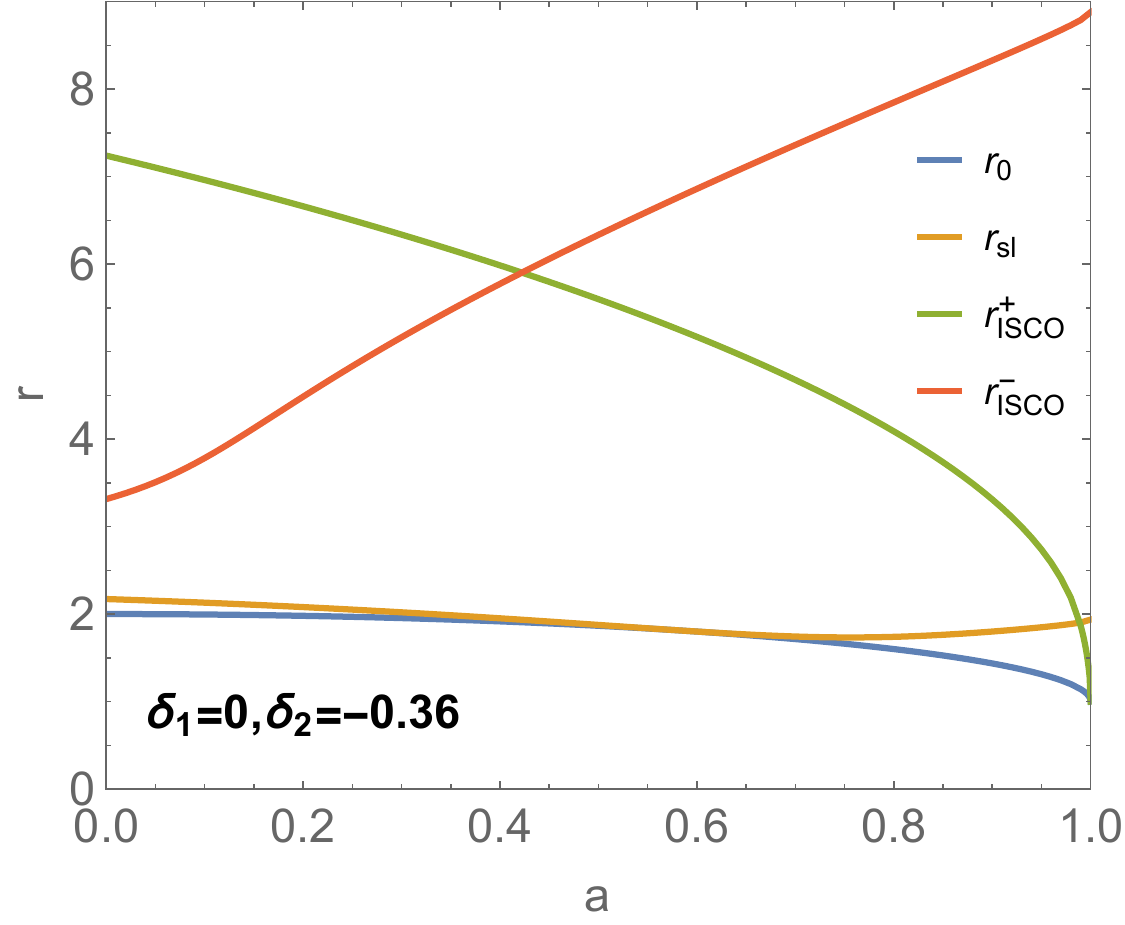}\quad
	\includegraphics[width=0.45\textwidth]{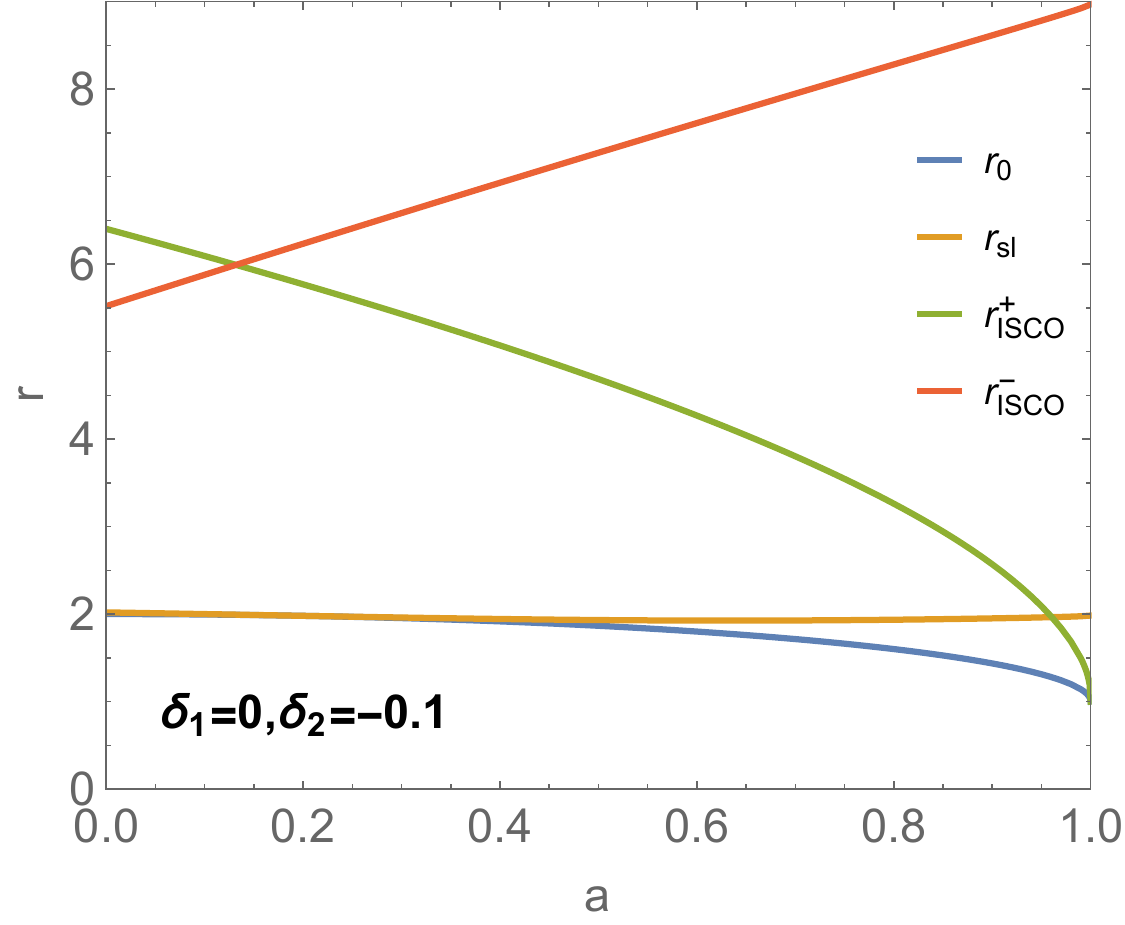}\\
	\vspace{0.5cm}
	\includegraphics[width=0.45\textwidth]{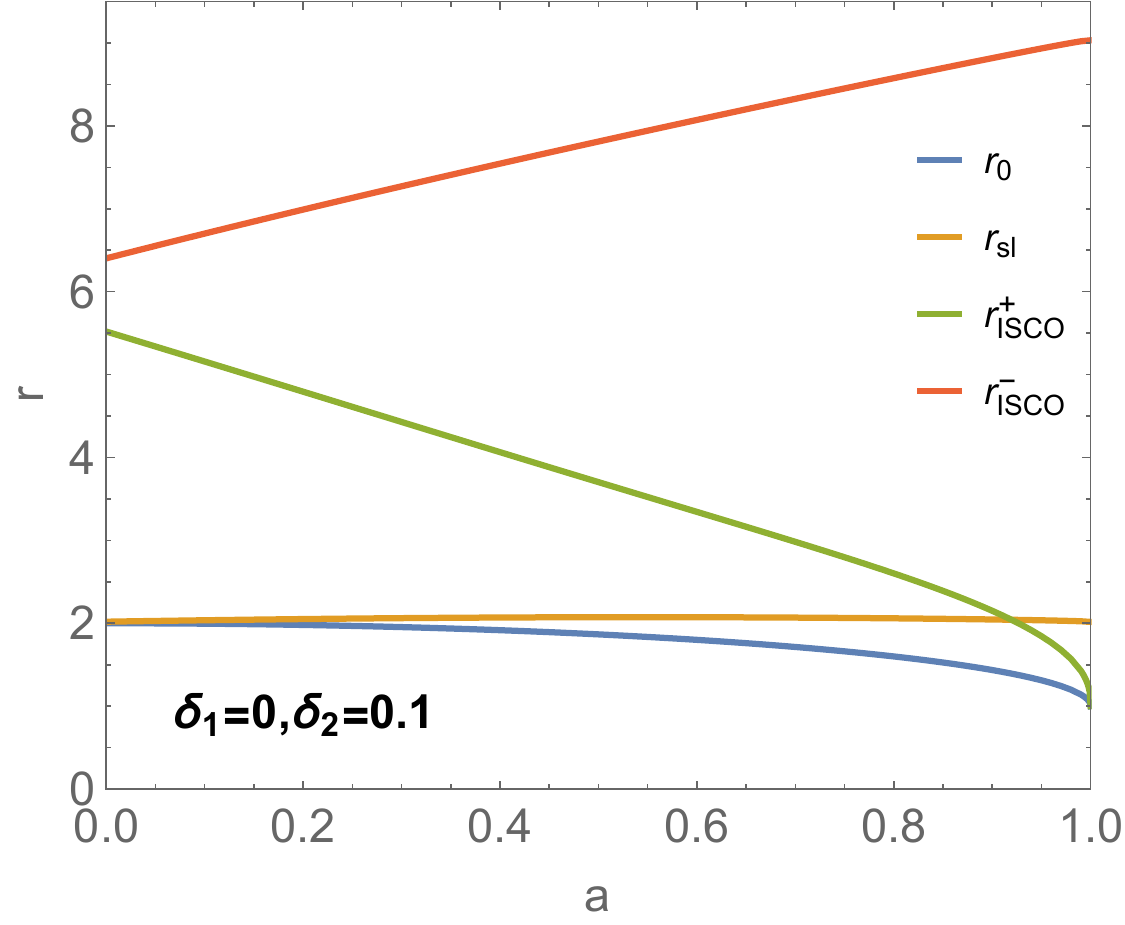}\quad
	\includegraphics[width=0.45\textwidth]{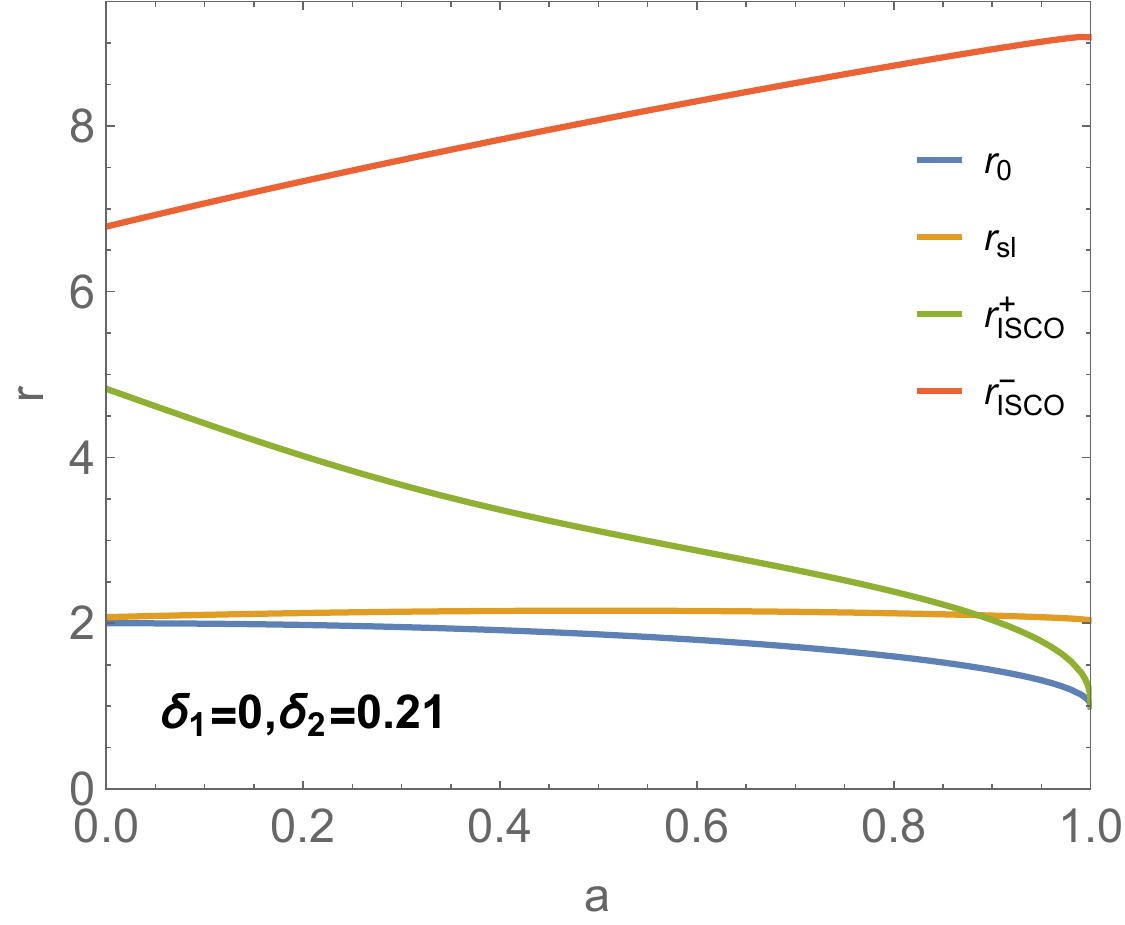}
	\caption{Radius of innermost stable circular orbit as a function of the spin $a$. Here we fix $\delta_1 =0$ and study the influence of $\delta_2$.  $r^+_{\rm ISCO}$ and $r^-_{\rm ISCO}$ represent the corotating and counter-rotating orbits, respectively. $r_0$ is the event horizon and $r_{sl}$ is the ergosphere.} \label{ISCODelta2}
\end{figure}

In Figs. \ref{ISCODelta1} and \ref{ISCODelta2} we plot the radius of the ISCO as a function of the spin $a$ for several typical values of $\{\delta_1, \delta_2\}$. The Kerr case is also shown for comparison. In each figure, we set one of the two parameters $\{\delta_1, \delta_2\}$ to zero and study the influence of the other. From the figures, we can see that as the spin $a$ increases, the radius of the counter-rotating ISCO $r^-_{\rm ISCO}$ increases monotonically and is always outside the ergoregion. Therefore, we will focus only on the corotating orbits. As the spin $a$ is increased, the radius of the corotating ISCO $r^+_{\rm ISCO}$ decreases almost monotonically. The only exception is when $\delta_1<0$ and $a$ is close to its extreme value $a=1$, where $r^+_{\rm ISCO}$ increases with increasing $a$. The corotating ISCO will enter the ergoregion when the spin $a$ is high enough to exceed some critical value $a_c$. From the figures, one can see that $\delta_1$ has little effect on $a_c$, while $\delta_2$ has a significant effect on $a_c$. As $|\delta_2|$ is increased, $a_c$ increases for $\delta_2 <0$ and decreases for $\delta_2 >0$. Since magnetic reconnection is related to orbits within the ergoregion, these results imply that for larger $|\delta_2|$, magnetic reconnection can occur for lower black hole spin if $\delta_2>0$, while for $\delta_2<0$ higher spin is required for magnetic reconnection to occur. Furthermore, $r^+_{\rm ISCO}$ does not coincide with the event horizon $r_0$ in the extreme limit when $\delta _1 \neq 0$, and $r^+_{\rm ISCO} \neq r^-_{\rm ISCO}$ in the nonrotating limit $a=0$ if $\delta _2 \neq 0$.

We have also checked that the vertically stable condition $\partial_\theta^2 V_{\rm eff} \leq 0$ is always satisfied for (radially) ISCO in the parameter region (\ref{ParameterRegion}).

\section{Energy extraction via magnetic reconnection}

In this section, we investigate the process of harnessing energy from the black hole through magnetic reconnection in the ergoregion, a phenomenon anticipated to be a common occurrence in rapidly spinning black holes. To simplify the analysis, we employ the methodology elucidated in \cite{Comisso:2020ykg} and focus on magnetic reconnection in the surrounding plasma, which rotates around the black hole in a stable circular orbit in the equatorial plane at the Keplerian velocity $\Omega_K$.

To assess the energy density of the outflowing plasma, it is advantageous to evaluate some quantities in the so-called ``zero-angular-momentum-observer'' (ZAMO) frame \cite{Bardeen:1972fi}. This is a locally nonrotating frame in which the spacetime is locally Minkowskian
\begin{align}
	ds^2 = - d \hat{t}^2 + \sum_{i=1}^3 (d \hat{x}^i)^2 = \eta_{\mu\nu} d\hat{x}^\mu d\hat{x}^\nu, 
\end{align} 
where the transformation between the ZAMO frame $(\hat{t}, \hat{x}^1 = \hat{r}, \hat{x}^2 = \hat{\theta}, \hat{x}^3 = \hat{\phi})$ and the Boyer-Lindquist coordinates $(t, x^1 = r, x^2 = \theta, x^3 = \phi)$ is
\begin{align}
	d \hat{t} = \alpha dt,\quad d\hat{x}^i = \sqrt{g_{ii}} dx^i - \alpha \beta^i dt,
\end{align} 
with the lapse function $\alpha$ and the shift vector $\beta^i = (0,0,\beta^\phi)$ being
\begin{align}
	\alpha = \left(- g_{tt} + \frac{g_{t\phi}^2}{g_{\phi\phi}}\right)^{1/2},\quad \beta^\phi = \frac{\sqrt{g_{\phi\phi}} \omega^\phi}{\alpha},
\end{align}
and $\omega^\phi = - g_{t\phi} / g_{\phi\phi}$ is the angular velocity of the frame dragging. We denote quantities in ZAMO frame with hats. The Keplerian velocity of the corotating bulk plasma in ZAMO becomes \begin{align}
	\hat{v}_K = \frac{d\hat{x}^\phi}{d\hat{t}} = \frac{\sqrt{g_{\phi\phi}}}{\alpha} \Omega_K - \beta^\phi.
\end{align}

Using the one-fluid approximation, the energy-momentum tensor of the plasma can be expressed as
\begin{align}
	T^{\mu\nu} = p g^{\mu\nu} + \omega U^\mu U^\nu + F^\mu_{~~\sigma} F^{\nu \sigma} - \frac{1}{4} g^{\mu\nu} F^{\rho\sigma} F_{\rho \sigma}, \label{EnergyMomentum}
\end{align} 
where $p, \omega, U^\mu$ and $F^{\mu\nu}$ are the proper plasma pressure, enthalpy density, four-velocity, and electromagnetic field tensors, respectively. With the timelike killing vector $\xi = \partial_t$, one can define a covariant conserved energy current $J^\mu \equiv T^{\mu \nu} \xi _\nu$ and the associated energy density $e^{\infty} \equiv n_\mu J^\mu = - \alpha g_{\mu 0} T^{\mu 0}$, where $n^\mu$ is the unit vector normal to timelike hypersurfaces $t=consant$. At infinity, $e^{\infty}$ is the energy density measured by static observers, and so is usually called the ``energy-at-infinity'' density. Assuming a highly efficient conversion of magnetic energy into kinetic energy, so that the electromagnetic component of the energy density can be neglected, and adopting the approximation that the plasma is incompressible and adiabatic, the energy-at-infinity density can be expressed as \cite{Comisso:2020ykg}
\begin{align}
	e^\infty = \alpha \hat{\gamma } \omega (1 + \beta^\phi \hat{v}^\phi) - \frac{\alpha p}{\hat{\gamma }},
\end{align}
where $\hat{\gamma} = \hat{U}^0$ is the Lorentz factor and $\hat{v}^\phi = d\hat{\phi} / d\hat{t}$ is the azimuthal component of the velocity of the outflow plasma. By considering both the accelerated and decelerated parts of the outflowing plasma, the energy-at-infinity density per enthalpy can ultimately be formulated as \cite{Comisso:2020ykg}
\begin{align}
	\epsilon^\infty_\pm = \alpha \hat{\gamma}_K \left[(1+\beta^\phi \hat{v}_K) (1+\sigma_0)^{1/2} \pm \cos\xi(\beta^\phi + \hat{v}_K) \sigma_0^{1/2} - \frac{1}{4} 
\frac{(1+\sigma_0)^{1/2} \mp \cos\xi \hat{v}_K \sigma_0^{1/2}}{\hat{\gamma}^2_K (1+\sigma_0 - \cos^2\xi \hat{v}^2_K \sigma_0)}\right], \label{EnergyAtInfinity}
\end{align}
where $\xi$ is the orientation angle between the outflow plasma velocity and the azimuthal direction in the equatorial plane in the local rest frame, $\hat{\gamma}_K= (1- \hat{v}_K^2)^{-1 /2}$ and $\sigma _0$ is the plasma magnetization. The signs $+$ and $-$ stand for accelerated and decelerated parts, respectively. From the above equation, it can be seen that the energy-at-infinity density is parametrized by five parameters $\{a, \delta _1, \delta _2, \sigma _0, \xi\}$ as well as the radius $r=r_X$ of the circular orbit of the plasma where the reconnection occurs (X-point). 

\begin{figure}[!htbp]
	\includegraphics[width=0.45\textwidth]{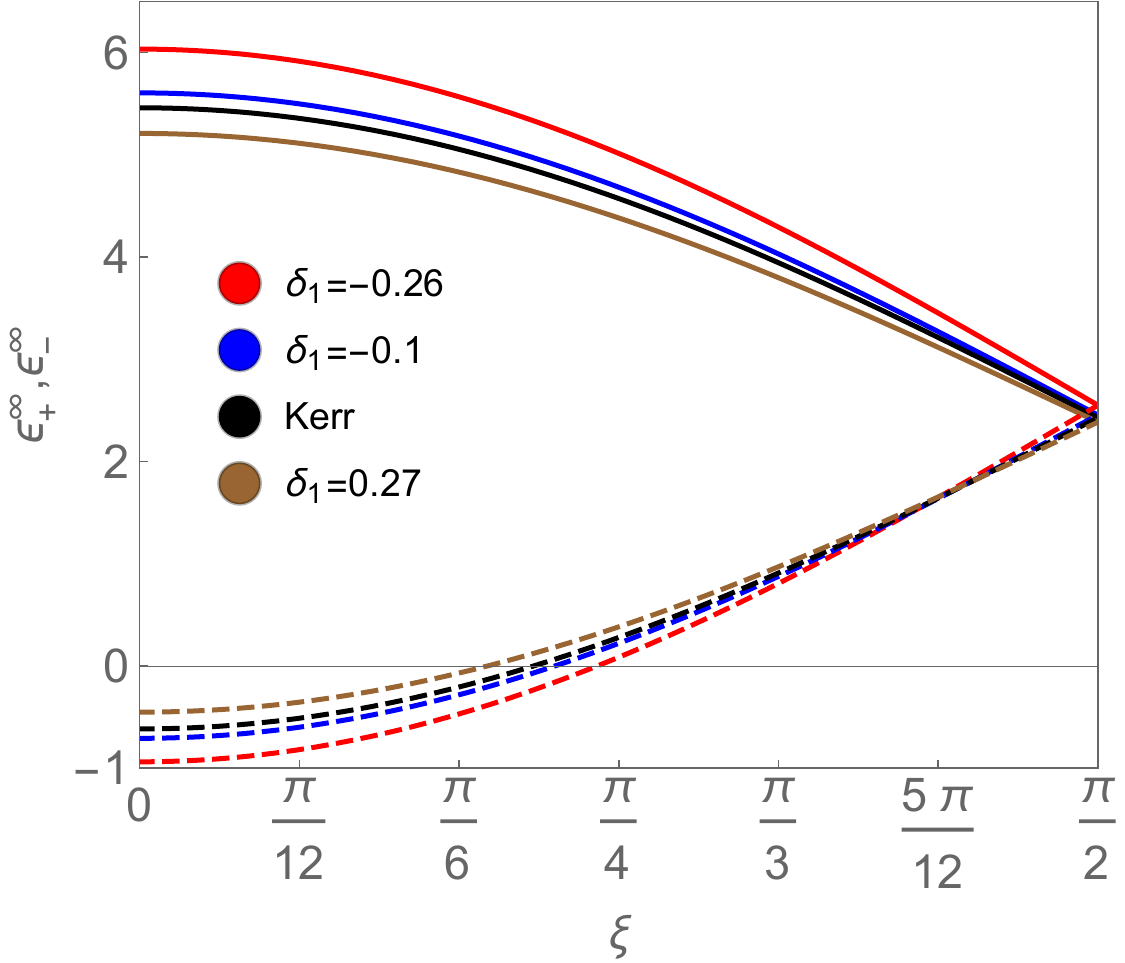}\quad
	\includegraphics[width=0.45\textwidth]{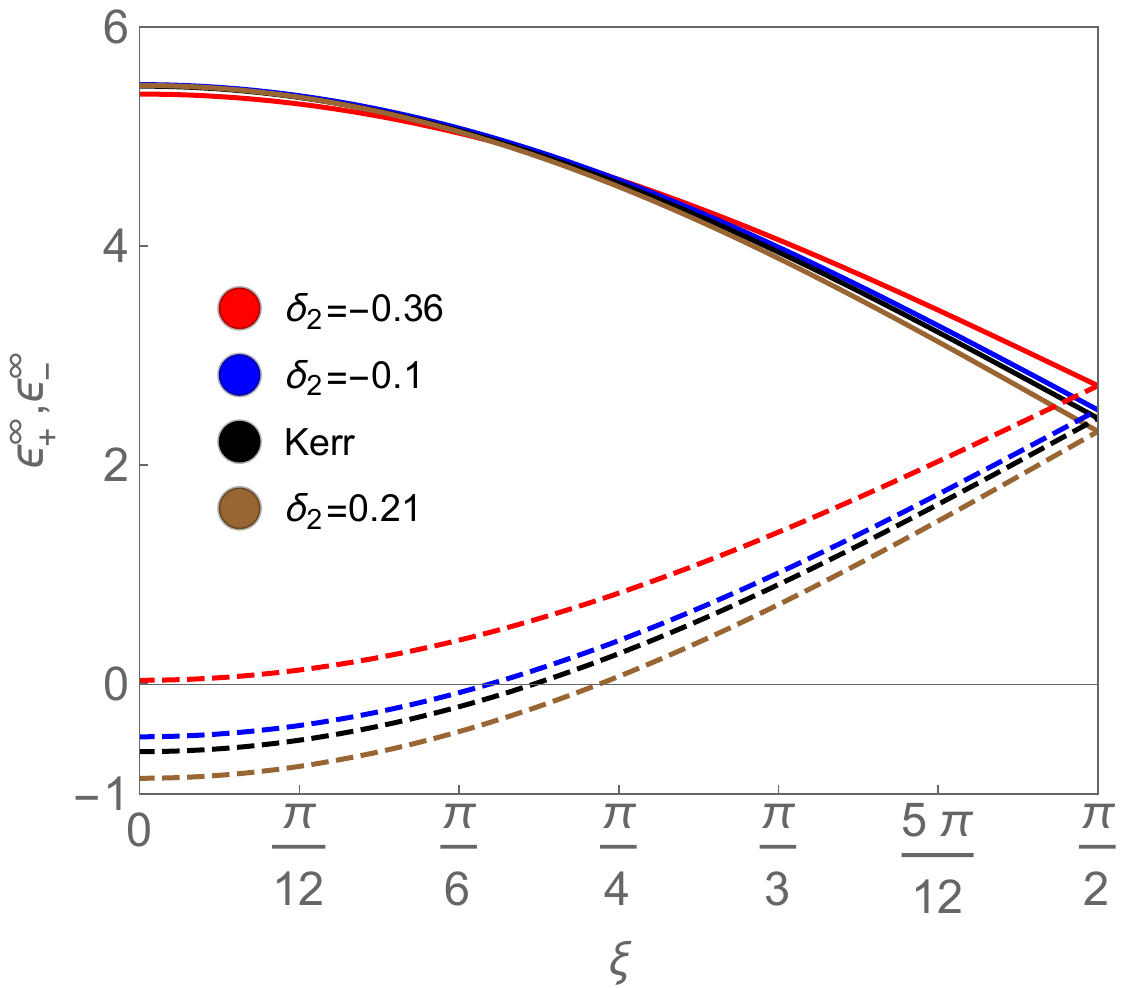}\\
	\vspace{0.5cm}
	\includegraphics[width=0.45\textwidth]{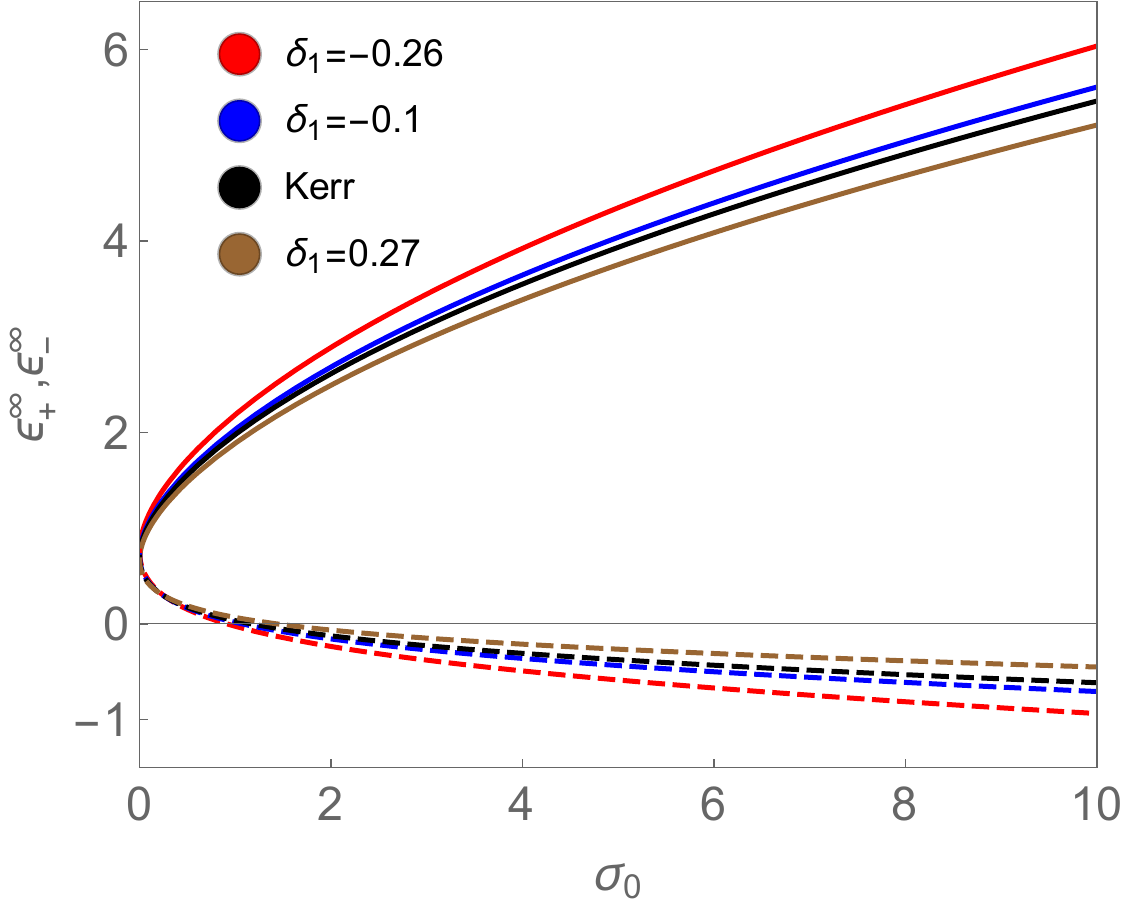}\quad
	\includegraphics[width=0.45\textwidth]{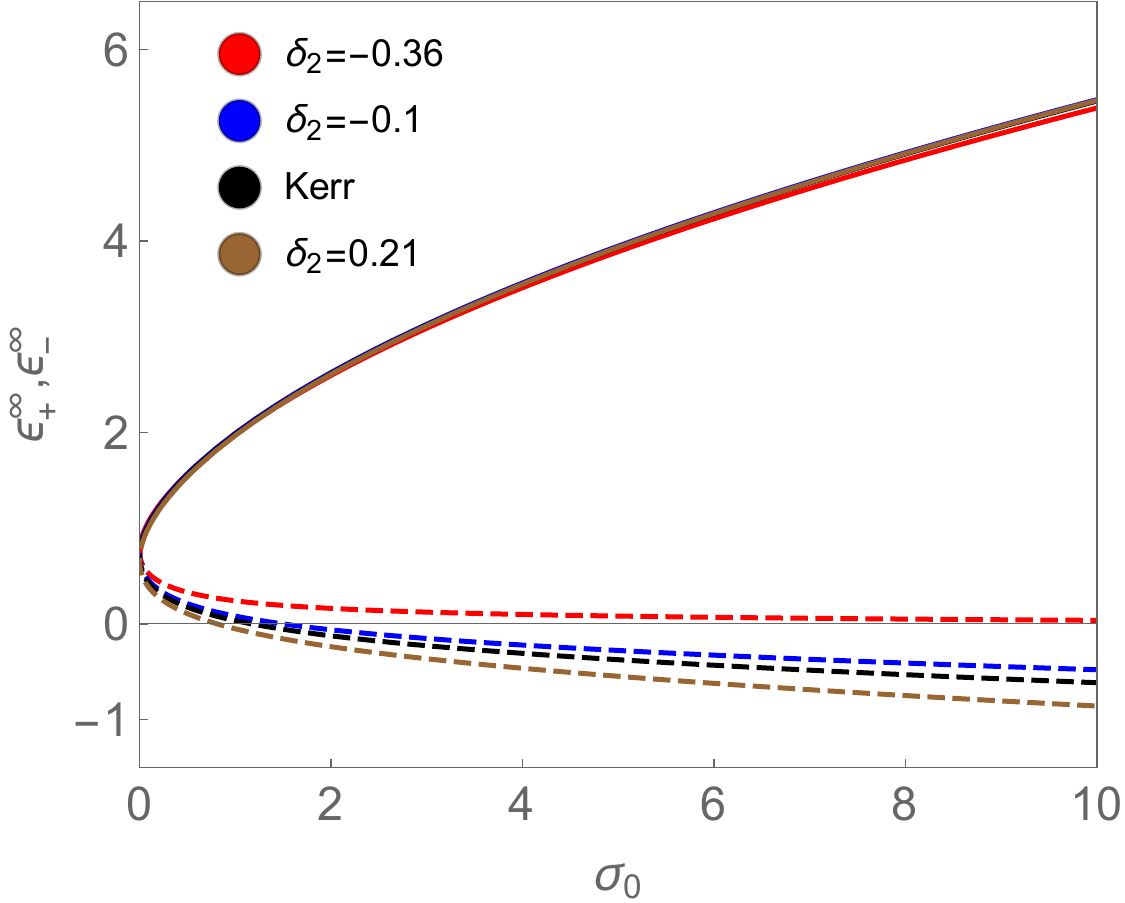}
	\caption{The energy-at-infinity density $\epsilon^\infty_\pm$ as a function of the orientation angle $\xi$ (top panels) and the plasma magnetization $\sigma _0$ (bottom panels) for various $\delta_1$ and $\delta_2$, with $a=0.99$ and $r_X = r^+_{\rm ISCO}$. In the top panels, we set $\sigma _0 =10$, while in the bottom panels, we set $\xi =0$. In the left panels $\delta_2=0$, while in the right panels $\delta_1=0$. Solid curves are $\epsilon ^\infty_+$ while dashed ones are $\epsilon ^\infty_-$. } \label{EpsilonXiSigma}
\end{figure}

Energy extraction from a black hole involves the requirement that decelerated plasma possesses negative energy-at-infinity, while accelerated plasma exhibits positive energy-at-infinity exceeding its rest mass and thermal energy, that is 
\begin{align}\label{EnergyExtractionConditions}
	\epsilon ^\infty_- <0, \quad \Delta \epsilon ^\infty_+ = \epsilon ^\infty_+ - \left(1 - \frac{\Gamma}{\Gamma -1} \frac{p}{\omega}\right) = \epsilon ^\infty_+ >0.
\end{align} 
Here we have assumed the plasma to be relativistically hot with polytropic index $\Gamma = 4 / 3$.

\subsection{Parameter space analysis}

In this subsection, we will perform a parameter space analysis to show the influences of $\{\delta _1, \delta _2\} $ on the permissible regions where the energy extraction conditions (\ref{EnergyExtractionConditions}) are satisfied.

In Fig. \ref{EpsilonXiSigma} we show the effects of $\{\delta _1, \delta _2\} $ on the required orientation angle $\xi$ and the plasma magnetization $\sigma _0$ to satisfy the energy extraction conditions (\ref{EnergyExtractionConditions}) for a near-extreme black hole ($a=0.99$). We take the X-point to be at $r_X = r^+_{\rm ISCO}$. From the figure we can see that, as in the Kerr case, energy extraction is favored by lower values of $\xi$ but higher values of $\sigma _0$. Moreover, $\epsilon ^\infty_+ > 0$ is always satisfied, while $\epsilon ^\infty_- < 0$ is only satisfied when $\xi$ is less than some upper bound $\xi ^c$ and $\sigma _0$ exceeds some lower bound $\sigma ^c _0$. As $\delta _1$ increases from negative to positive values, $\xi ^c$ is shifted to smaller values, and $\sigma _0^c$ is shifted to larger values, while $\delta _2$ has the opposite effect. Furthermore, if $\delta _2$ is negative enough, $\epsilon ^\infty_- < 0$ cannot be satisfied for any $\xi$ and $\sigma _0$. Compared to the Kerr case, smaller $\sigma_0^c$ is required to satisfy the energy extraction conditions when $\delta_2<0$ or $\delta_1>0$. These results imply that energy extraction is favored by negative $\delta _1$ but positive $\delta _2$.

\begin{figure}[!htbp]
	\includegraphics[width=0.45\textwidth]{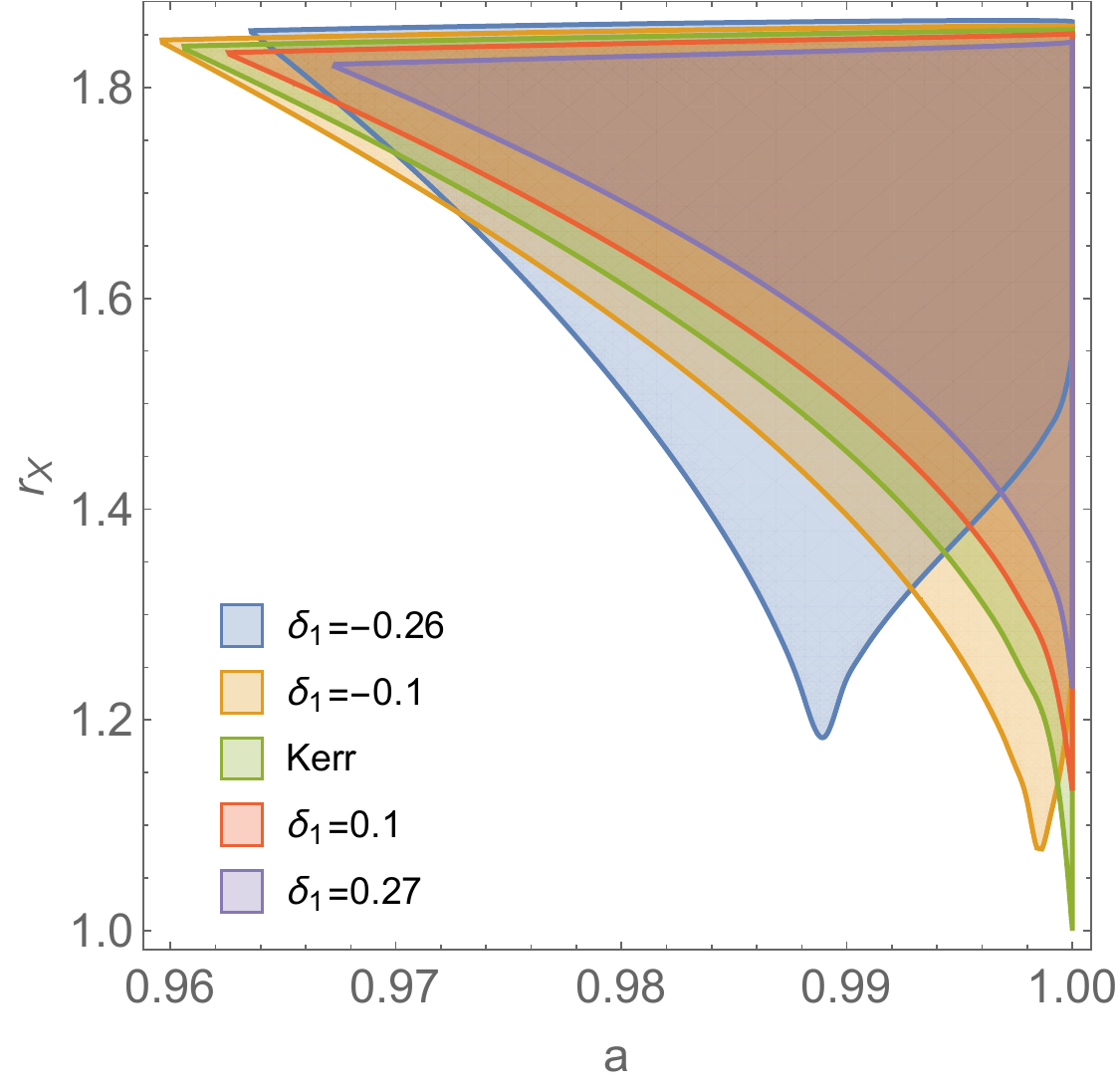}\quad
	\includegraphics[width=0.45\textwidth]{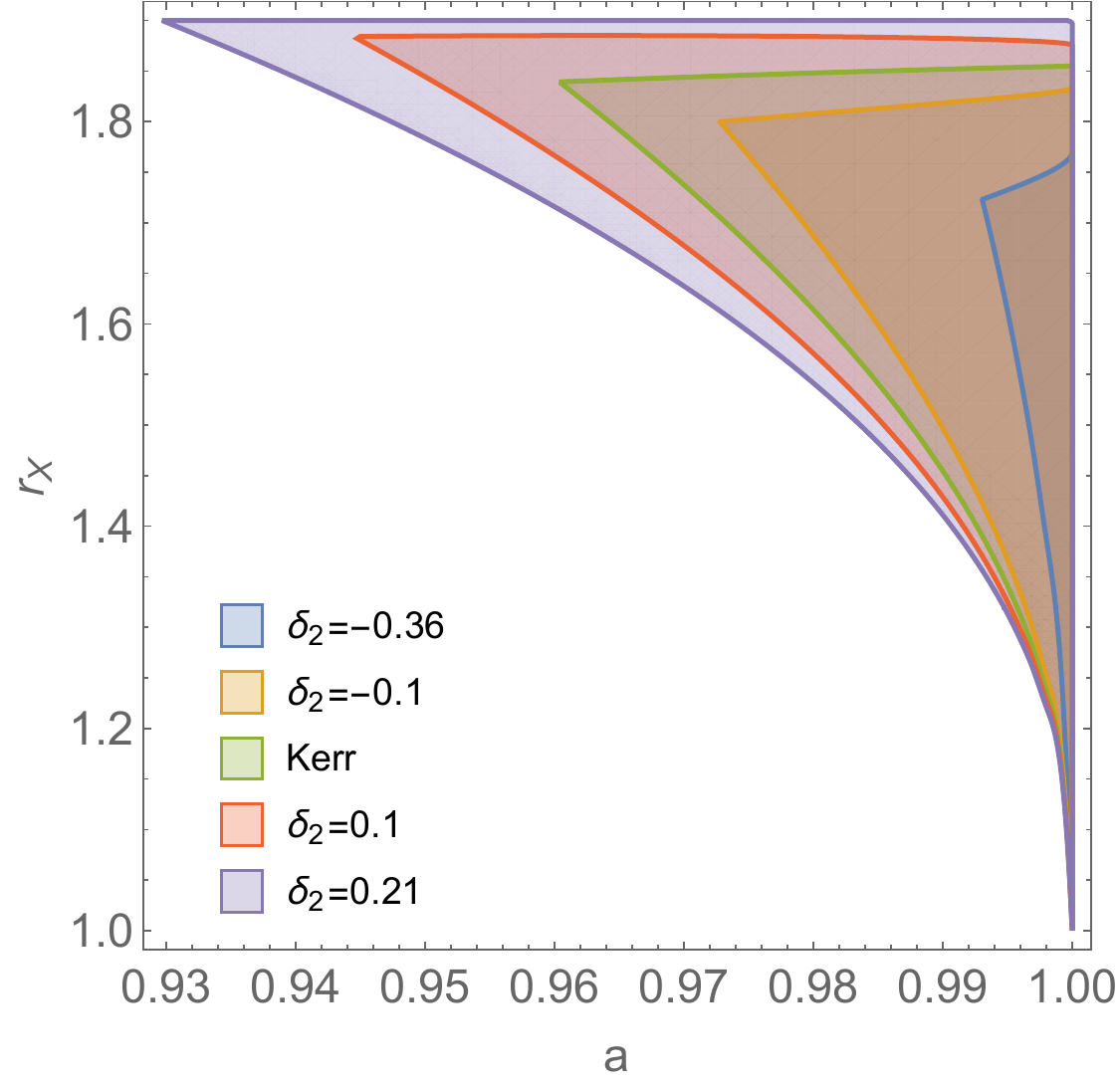}
	\caption{Permissible regions (shaded) in $a-r_X$ plane where the energy extraction conditions are satisfied for $r^+_{\rm ISCO}\leq r_X<r_{\rm sl}$. We set $\xi=\frac{\pi}{12}$ and $\sigma _0=10$. In the left panels $\delta_2=0$, while in the right panels $\delta_1=0$.} \label{Epsilonra}
\end{figure}

In Fig. \ref{Epsilonra} we plot the permissible regions in the $a-r_X$ plane where the energy extraction conditions (\ref{EnergyExtractionConditions}) are satisfied for various $\delta _1$ and $\delta _2$. Note that the bulk plasma is assumed to rotate in a stable circular orbit in the ergoregion, so the permissible regions are additionally constrained by $r^+_{\rm ISCO}\leq r_X<r_{\rm sl}$. Numerically, we found that the lower boundary of the permissible region is given by $r_X=r^+_{\rm ISCO}$, while the upper boundary is inside the ergosphere. Thus, $r_X<r_{\rm sl}$ is not sufficient to satisfy the energy extraction conditions. Only when the X-point is deep enough in the ergoregion can the energy extraction conditions be satisfied. From the right panel, it can be seen that as $\delta _2$ decreases from a positive value to a negative value, the permissible region shrinks significantly and a higher $a$ is required to satisfy the energy extraction conditions. It is worth noting that, for $\delta_2=0.21$, $a \gtrsim 0.93$ is necessary to fulfill the energy extraction conditions. Conversely, for $\delta_2=-0.36$, an exceedingly high spin of $a \gtrsim 0.994$ is needed to meet the energy extraction conditions. For comparison, in the Kerr case, $a \gtrsim 0.96$ is necessary to fulfill the energy extraction conditions. This again confirms our previous conclusion that energy extraction is favored by positive $\delta _2$. However, the influence of $\delta _1$ on the permissible region is not so clear as shown in the left panel. It can be seen that as $\delta _1$ increases from negative to positive, the required lower bound of $a$ first decreases and then increases moderately. This complication is due to the complicated effect of $\delta _1$ on $r^+_{\rm ISCO}$ as previously shown in Fig. \ref{ISCODelta1}.

\subsection{Energy extraction power and efficiency}

In this subsection, we analyze the effects of $\{\delta _1, \delta _2\} $ on the power and efficiency of energy extraction.

\begin{figure}[!htbp]
	\includegraphics[width=0.45\textwidth]{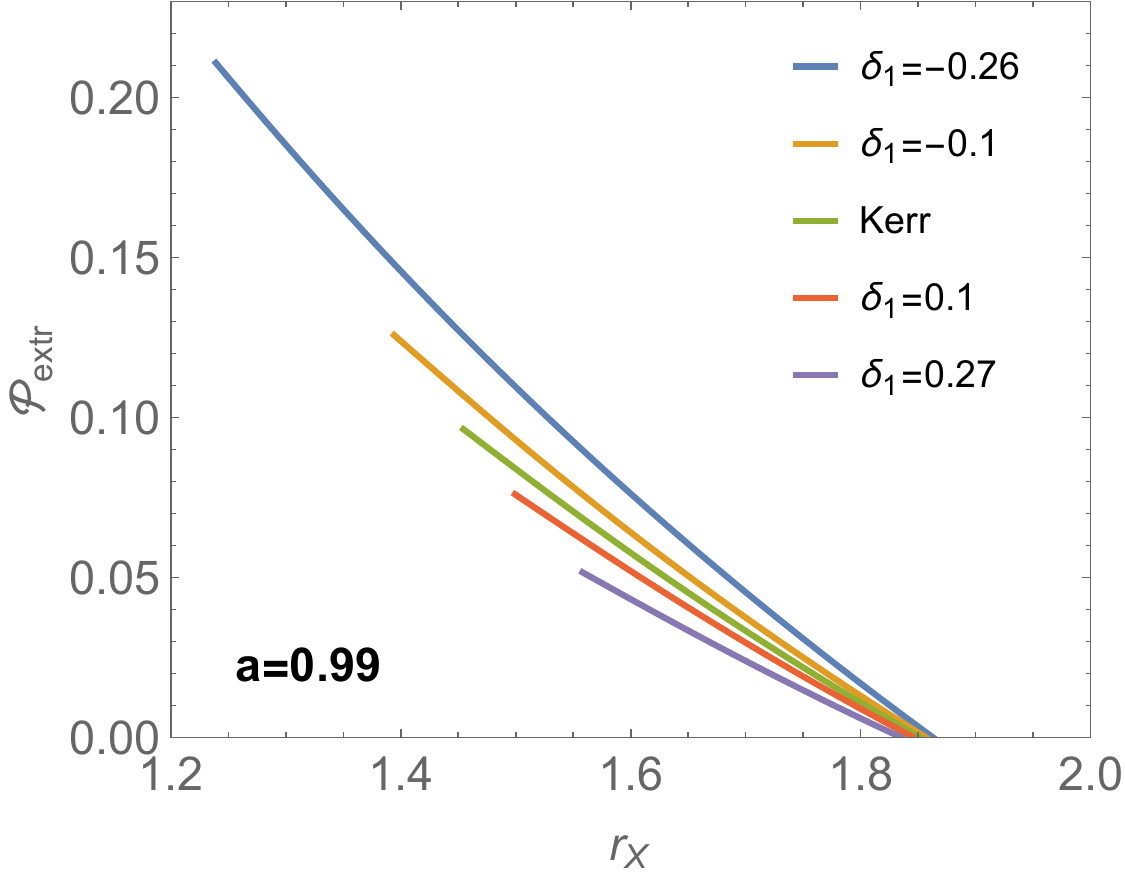}\quad
	\includegraphics[width=0.45\textwidth]{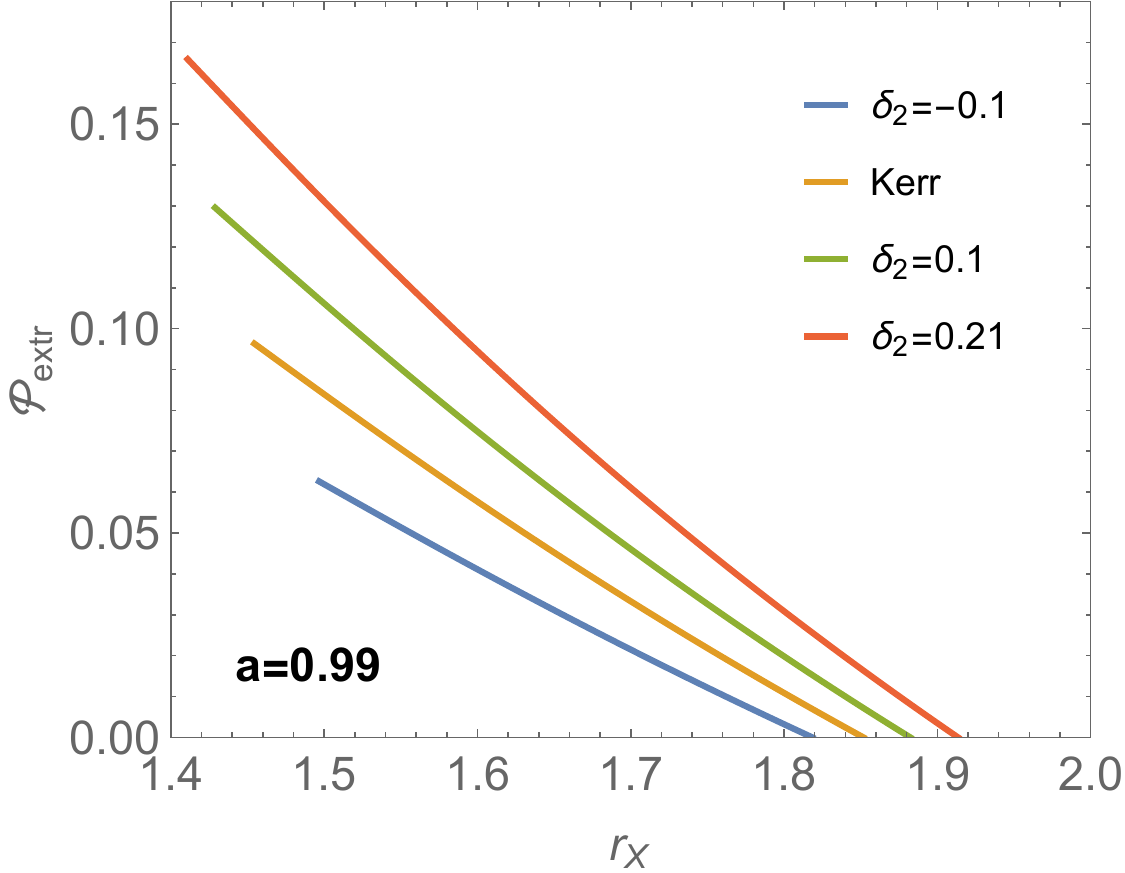}\\
	\vspace{0.5cm}
	\includegraphics[width=0.45\textwidth]{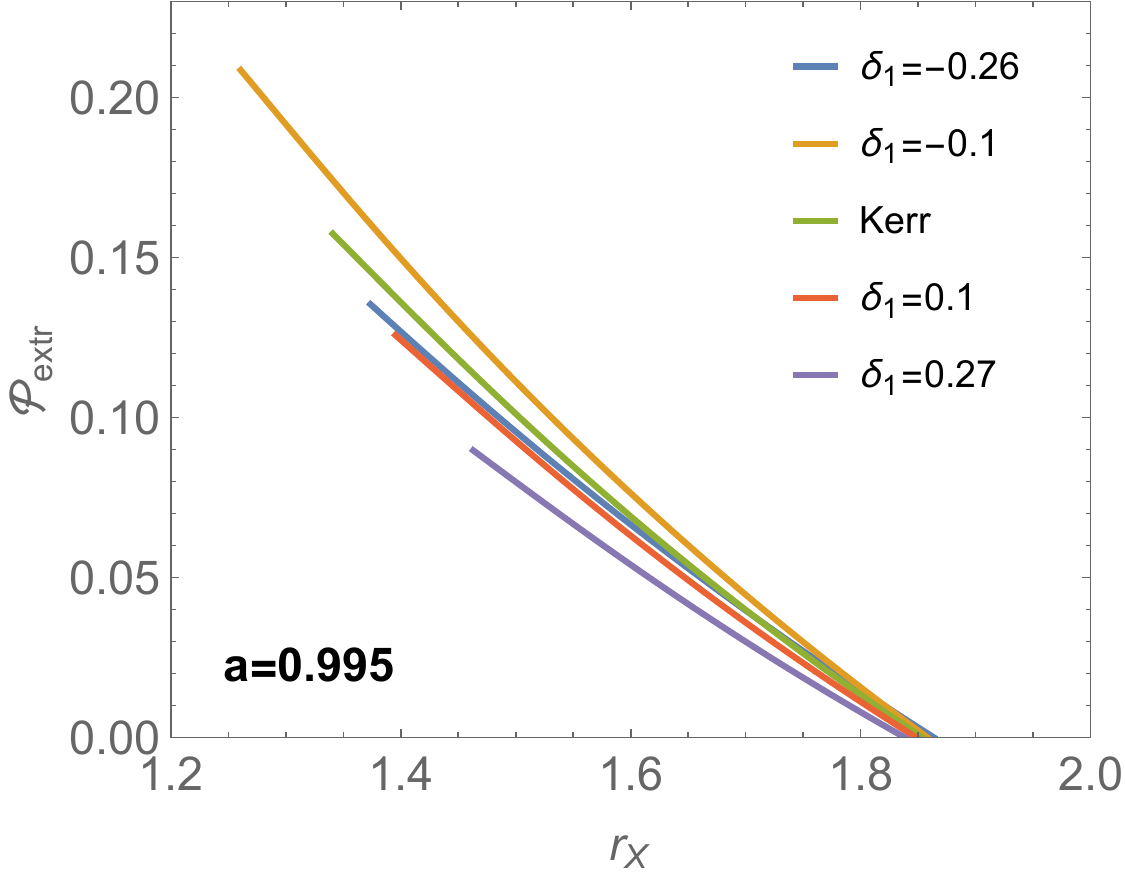}\quad
	\includegraphics[width=0.45\textwidth]{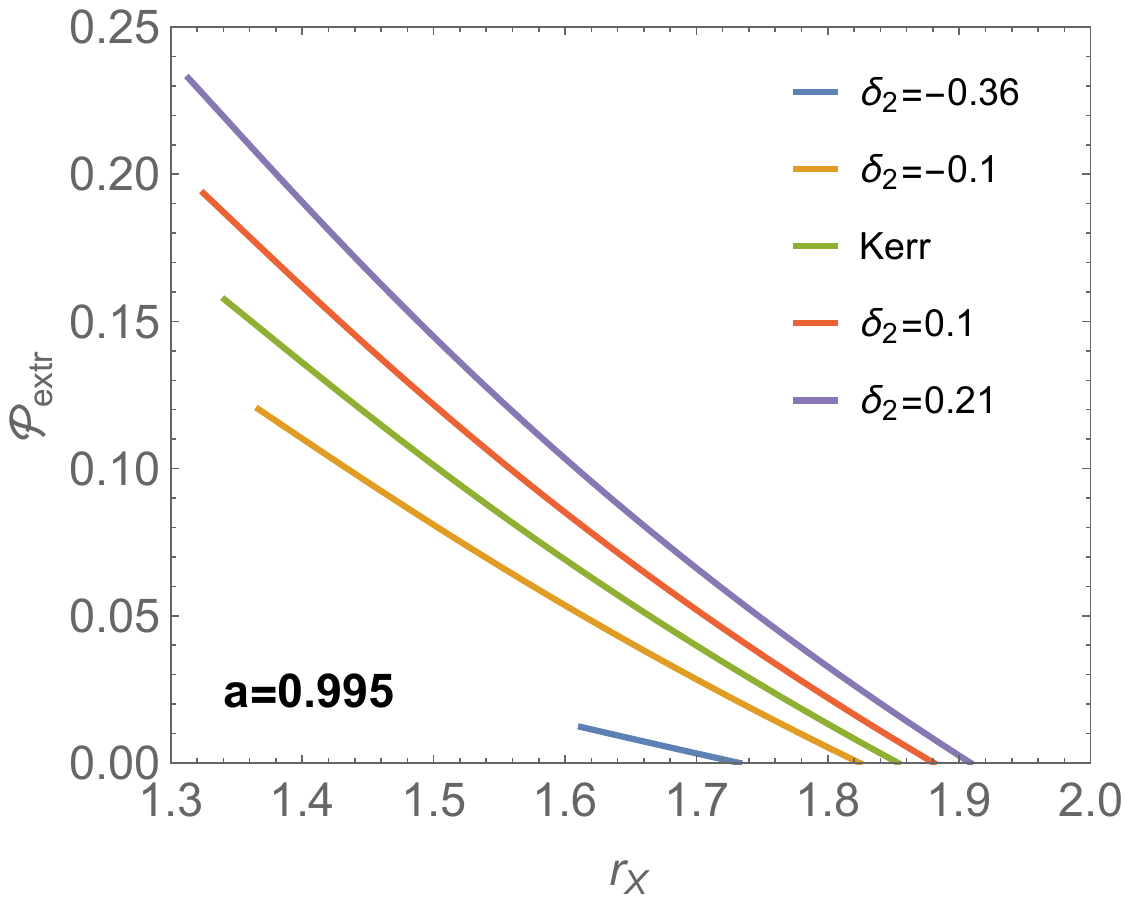}
	\caption{${\cal P}_{\rm extr}$ as a function of the X-point location $r_X$ for various $\{\delta _1, \delta _2\}$, with $\xi=\frac{\pi}{12},\sigma _0=10$ and $U_{\rm in}=0.1$. $r_X$ is restricted to be in the range $r^+_{\rm ISCO}\leq r_X<r_{\rm sl}$. In the left panels $\delta_2=0$, while in the right panels $\delta_1=0$.} \label{PowerR}
\end{figure}

\begin{figure}[!htbp]
	\includegraphics[width=0.45\textwidth]{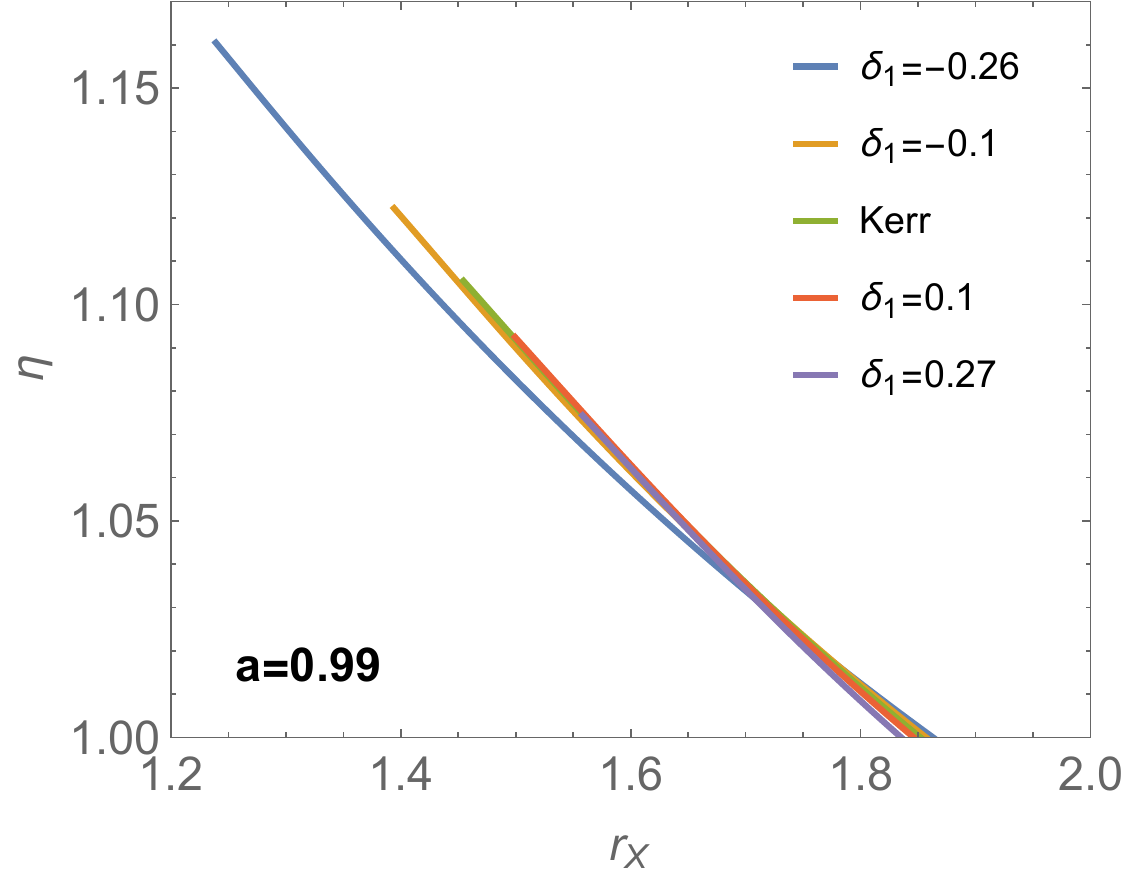}\quad
	\includegraphics[width=0.45\textwidth]{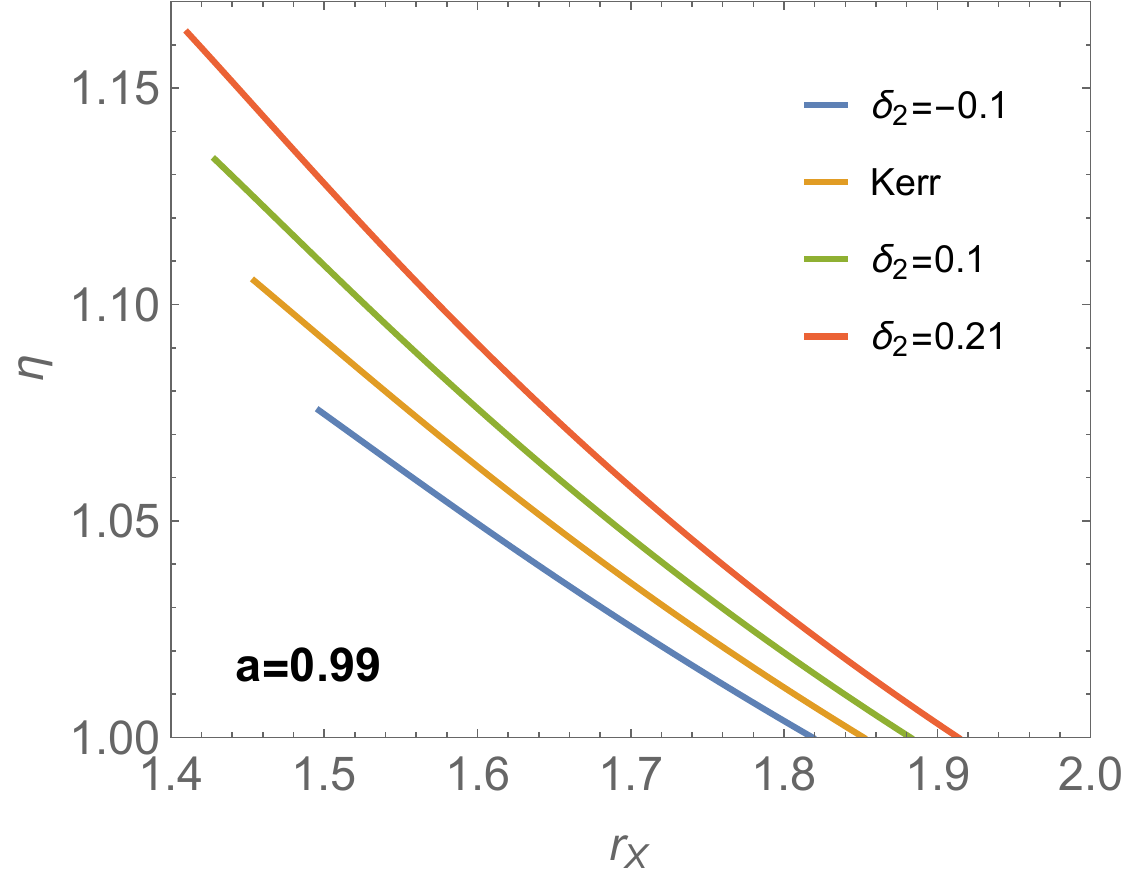}\\
	\vspace{0.5cm}
	\includegraphics[width=0.45\textwidth]{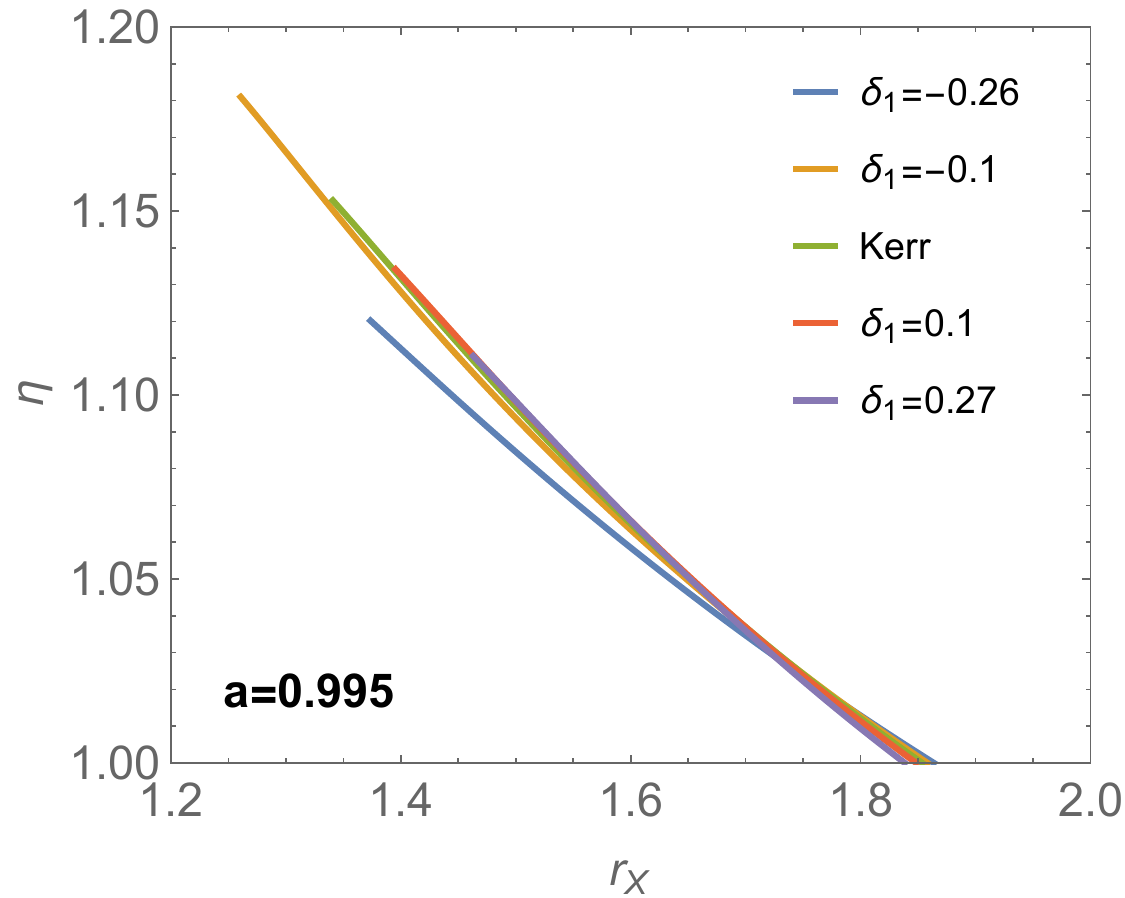}\quad
	\includegraphics[width=0.45\textwidth]{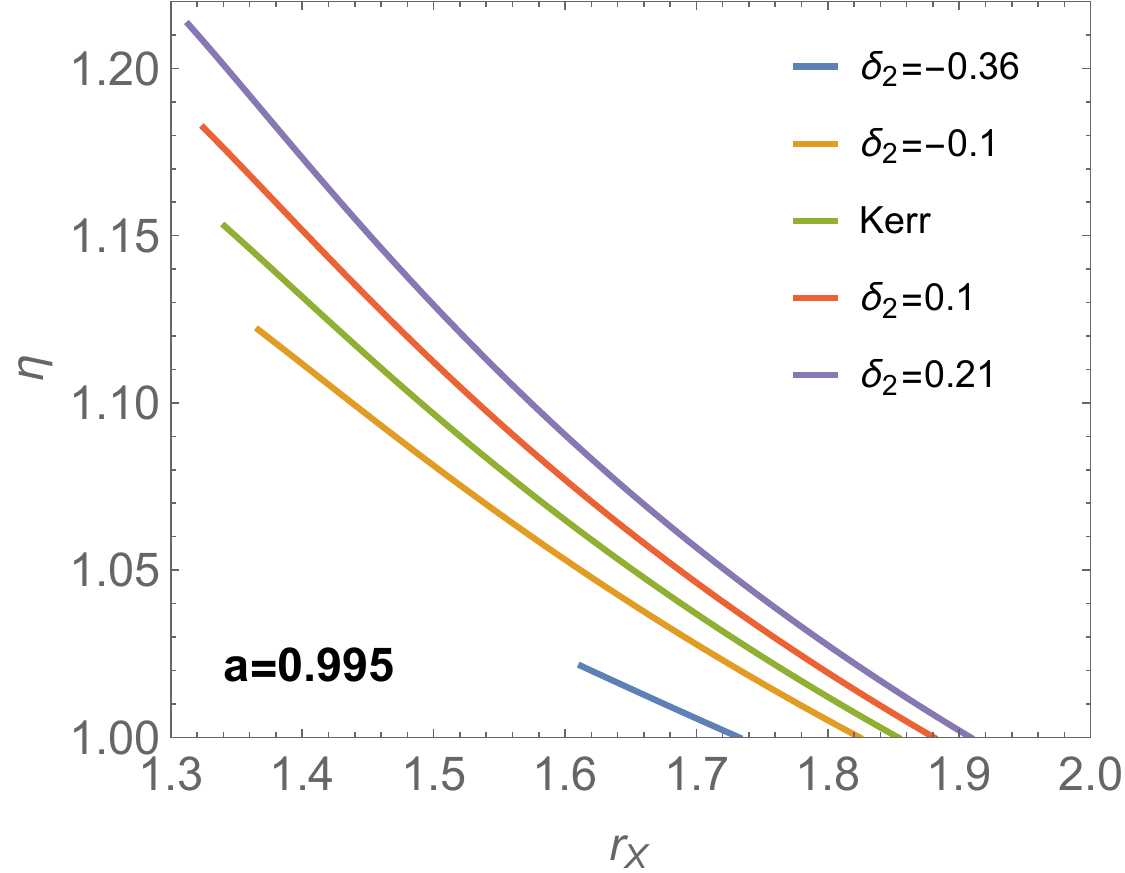}
	\caption{$\eta$ as a function of the X-point location $r_X$ for various $\{\delta _1, \delta _2\}$, with $\xi=\frac{\pi}{12},\sigma _0=10$ and $U_{\rm in}=0.1$. $r_X$ is restricted to be in the range $r^+_{\rm ISCO}\leq r_X<r_{\rm sl}$. In the left panels $\delta_2=0$, while in the right panels $\delta_1=0$.} \label{EfficiencyR}
\end{figure}

The power ${\cal P}_{\rm extr}$ per unit enthalpy extracted from the black hole by the escaping plasma can be well-estimated as \cite{Comisso:2020ykg}
\begin{align}
	{\cal P}_{\rm extr} = - \epsilon ^\infty_- A_{\rm in} U_{\rm in},
\end{align} 
where $A_{\rm in}$ is the cross sectional area of the inflowing plasma, which can be estimated as $A_{\rm in} \sim (r_{\rm sl}^2 - r_{\rm ISCO}^2)$ for fast-rotating black holes. The parameter $U_{\rm in} = {\cal O}(10^{-2})$ and ${\cal O}(10^{-1})$ for the collisional and collisionless regime respectively.

Figure \ref{PowerR} shows a typical picture of power ${\cal P}_{\rm extr}$ as a function of the X-point location $r_X$. From the figure it can be seen that, with all other parameters fixed, ${\cal P}_{\rm extr}$ is a monotonically decreasing function of $r_X$, reaching a maximum at $r_X=r^+_{\rm ISCO}$. From the left panels, we can see that for fixed $r_X$ the effect of $\delta_1$ on ${\cal P}_{\rm extr}$ is moderate and depends on the spin $a$. When $a$ is not so extremely high, ${\cal P}_{\rm extr}$ is a monotonically decreasing function of $\delta_1$. However, when $a$ is extremely high, as $\delta_1$ increases, ${\cal P}_{\rm extr}$ first increases and then decreases. From the right panels, we can see that for fixed $r_X$, ${\cal P}_{\rm extr}$ is a monotonically increasing function of $\delta_2$. Note that for $a<0.994$, as we mentioned above, there is no energy extraction for $\delta_2 = -0.36$. For given $a$, the maximum power for the KRZ black hole can significantly exceed that of the Kerr case when $\{\delta_1, \delta_2\}$ fall within particular ranges. For instance, when $a=0.99$, $\frac{{\cal P}^{\rm max}_{\rm extr} (\delta_1 =-0.26, \delta_2 =0)}{{\cal P}^{\rm max}_{\rm extr} ({\rm Kerr})} \sim 2.0$ and  $\frac{{\cal P}^{\rm max}_{\rm extr} (\delta_1 =0, \delta_2 =0.21)}{{\cal P}^{\rm max}_{\rm extr} ({\rm Kerr})} \sim 1.7$. Similarly, for $a=0.995$, $\frac{{\cal P}^{\rm max}_{\rm extr} (\delta_1 =-0.1, \delta_2 =0)}{{\cal P}^{\rm max}_{\rm extr} ({\rm Kerr})} \sim 1.5$ and  $\frac{{\cal P}^{\rm max}_{\rm extr} (\delta_1 =0, \delta_2 =0.21)}{{\cal P}^{\rm max}_{\rm extr} ({\rm Kerr})} \sim 1.5$.

To further measure the feasibility of magnetic reconnection energy extraction, it is convenient to define the efficiency of the plasma energization process \cite{Comisso:2020ykg},
\begin{align}
	\eta = \frac{\epsilon^\infty_+}{\epsilon^\infty_+ + \epsilon^\infty_-}.
\end{align}
If $\eta>1$, energy is extracted from the black hole. Figure \ref{EfficiencyR} shows a typical picture of the efficiency $\eta$ as a function of the X-point location $r_X$. From the figure it can be seen that, for other fixed parameters, $\eta$ is a monotonically decreasing function of $r_X$, reaching a maximum at $r_X=r^+_{\rm ISCO}$. The right panels show that for fixed $r_X$, $\eta$ is a monotonically increasing function of $\delta_2$. However, the effect of $\delta_1$ on $\eta$ is complicated and moderately dependent on 
$r_X$. When $r_X$ is small, $\eta$ is a monotonically increasing function of $\delta_1$. However, if $r_X$ is large enough, $\eta$ becomes a monotonically decreasing function of $\delta_1$. For given $a$, the maximum efficiency for the KRZ black hole can also exceed that of the Kerr case when $\{\delta_1, \delta_2\}$ fall within particular ranges. For instance, when $a=0.99$, $\frac{\eta^{\rm max} (\delta_1 =-0.26, \delta_2 =0)}{\eta^{\rm max} ({\rm Kerr})} \sim 105\%$ and  $\frac{\eta^{\rm max} (\delta_1 =0, \delta_2 =0.21)}{\eta^{\rm max} ({\rm Kerr})} \sim 106\%$. Similarly, for $a=0.995$, $\frac{\eta^{\rm max} (\delta_1 =-0.1, \delta_2 =0)}{\eta^{\rm max} ({\rm Kerr})} \sim 106\%$ and  $\frac{\eta^{\rm max} (\delta_1 =0, \delta_2 =0.21)}{\eta^{\rm max} ({\rm Kerr})} \sim 109\%$. 

\section{Summary and conclusions}

In this work, we study energy extraction from the KRZ parametrized black holes via the Comisso-Asenjo mechanism \cite{Comisso:2020ykg}, a novel mechanism recently proposed to extract energy from black holes via the magnetic reconnection process. The KRZ metric is a parametrized metric that describes generic stationary and axisymmetric black holes \cite{Konoplya:2016jvv}, with six parameters introduced to characterize deviations from the Kerr metric. Of these parameters, $\delta_1$ and $\delta_2$ play a crucial role in the physics under consideration, corresponding to deformations of $g_{tt}$ and rotational deformations of the metric, respectively, especially in the near-horizon zone. The values of both parameters are constrained by current astronomical observations (\ref{ParameterRegion}). We study in detail the ergoregion, the circular geodesics in the equatorial plane, and the energy extraction via magnetic reconnection, and find that these two deformation parameters have significant influences.

Magnetic reconnection takes place in the ergoregion. Following \cite{Koide:2008xr}, we have assumed that the bulk plasma rotates around the black hole in a circular and stable orbit in the equatorial plane. Such an orbit exists from infinity to ISCO. As shown in Figs. \ref{ErgoregionEquatorial}, \ref{ISCODelta1} and \ref{ISCODelta2}, $\{\delta_1, \delta_2\}$ significantly affects the shape of the ergoregion and the location of ISCO. $\{\delta_1, \delta_2\}$ can enlarge or shrink the ergoregion depending on their signs and the spin $a$. Specifically, a negative $\delta_1$ and a positive $\delta_2$ promote magnetic reconnection by enlarging the ergoregion. The counter-rotating ISCO $r^-_{\rm ISCO}$ is always outside the ergoregion and is therefore not relevant to the process. The corotating ISCO enters the ergoregion when the spin $a$ is high enough to exceed some critical value $a_c$. $\delta_1$ has little effect on $a_c$, while $\delta_2$ has a significant effect on $a_c$. A higher positive value of $\delta_2$ is advantageous for magnetic reconnection, as it necessitates a lower black hole spin for the process to occur. 

To realize energy extraction, the energy extraction conditions (\ref{EnergyExtractionConditions}), $\epsilon ^\infty_+ > 0$ and $\epsilon ^\infty_- < 0$, must be satisfied. We have performed a detailed analysis of the influences of the deformation parameters $\{\delta_1, \delta_2\}$ on the energy extraction conditions in Figs. \ref{EpsilonXiSigma}, \ref{Epsilonra}, \ref{PowerR} and \ref{EfficiencyR}. 

From Fig. \ref{EpsilonXiSigma} it can be seen that an increase in the magnitude of negative $\delta_1$ or positive $\delta_2$ leads to a reduction in the critical plasma magnetization $\sigma_0$ required for energy extraction. This observation indicates a preference for energy extraction with negative $\delta_1$ and positive $\delta_2$. From Fig. \ref{Epsilonra} it can be seen that as $\delta _2$ decreases from positive to negative, the permissible region in the $a-r_X$ plane to satisfy the energy extraction conditions shrinks significantly and a higher value of $a$ is required. It is noteworthy that if $\delta_2$ is sufficiently negative, an exceedingly high spin (e.g., $a \gtrsim 0.994$ for $\delta _2 = -0.36$) is required to satisfy the energy extraction conditions, posing challenges for energy extraction via this mechanism. This again confirms our previous conclusion that energy extraction is favored by a large positive $\delta _2$. This conclusion is further confirmed by examining the effects of the deformation parameters $\{\delta_1, \delta_2\}$ on the power ${\cal P}_{\rm extr}$ (Fig. \ref{PowerR}) and efficiency $\eta$ (Fig. \ref{EfficiencyR}). All these results imply that energy extraction is favored by negative $\delta _1$ but positive $\delta _2$. Compared to the Kerr case, the maximum power and efficiency for the KRZ black hole can be significantly higher.

Typically, the energy of the matter and electromagnetic fields surrounding a black hole is negligible compared to the mass of the black hole. Therefore, in our work, as in the related studies mentioned earlier, we have not considered the backreactions of the surrounding matter and electromagnetic fields on the black hole geometry.  Non-Kerr rotating black holes typically have only a small deviation from the Kerr black hole. If the energy of the matter and fields surrounding these black holes is sufficiently high, their backreaction on the geometry can be comparable to the deviation. In this case, it is necessary to consider their effects on the black hole geometry and the energy extraction process. Further studies are needed to fully understand these effects.

\begin{acknowledgments}

	This work is supported by the National Natural Science Foundation of China (NNSFC) under Grant No 12075207.

\end{acknowledgments}

\bibliographystyle{utphys}
\bibliography{Refslib}

\providecommand{\href}[2]{#2}\begingroup\raggedright\begin{thebibliography}{10}

\bibitem{LIGOScientific:2016aoc}
{\bfseries LIGO Scientific, Virgo} Collaboration, B.~P. Abbott {\em et~al.},
  ``{Observation of Gravitational Waves from a Binary Black Hole Merger},''
  \href{https://dx.doi.org/10.1103/PhysRevLett.116.061102}{{\em Phys. Rev.
  Lett.} {\bfseries 116} no.~6, (2016) 061102},
  \href{https://arxiv.org/abs/1602.03837}{{\ttfamily arXiv:1602.03837
  [gr-qc]}}.

\bibitem{LIGOScientific:2016sjg}
{\bfseries LIGO Scientific, Virgo} Collaboration, B.~P. Abbott {\em et~al.},
  ``{GW151226: Observation of Gravitational Waves from a 22-Solar-Mass Binary
  Black Hole Coalescence},''
  \href{https://dx.doi.org/10.1103/PhysRevLett.116.241103}{{\em Phys. Rev.
  Lett.} {\bfseries 116} no.~24, (2016) 241103},
  \href{https://arxiv.org/abs/1606.04855}{{\ttfamily arXiv:1606.04855
  [gr-qc]}}.

\bibitem{LIGOScientific:2017bnn}
{\bfseries LIGO Scientific, VIRGO} Collaboration, B.~P. Abbott {\em et~al.},
  ``{GW170104: Observation of a 50-Solar-Mass Binary Black Hole Coalescence at
  Redshift 0.2}'' \href{https://dx.doi.org/10.1103/PhysRevLett.118.221101}{{\em
  Phys. Rev. Lett.} {\bfseries 118} no.~22, (2017) 221101},
  \href{https://arxiv.org/abs/1706.01812}{{\ttfamily arXiv:1706.01812
  [gr-qc]}}. [Erratum: Phys.Rev.Lett. 121, 129901 (2018)].

\bibitem{LIGOScientific:2017ycc}
{\bfseries LIGO Scientific, Virgo} Collaboration, B.~P. Abbott {\em et~al.},
  ``{GW170814: A Three-Detector Observation of Gravitational Waves from a
  Binary Black Hole Coalescence},''
  \href{https://dx.doi.org/10.1103/PhysRevLett.119.141101}{{\em Phys. Rev.
  Lett.} {\bfseries 119} no.~14, (2017) 141101},
  \href{https://arxiv.org/abs/1709.09660}{{\ttfamily arXiv:1709.09660
  [gr-qc]}}.

\bibitem{LIGOScientific:2020iuh}
{\bfseries LIGO Scientific, Virgo} Collaboration, R.~Abbott {\em et~al.},
  ``{GW190521: A Binary Black Hole Merger with a Total Mass of $150
  M_{\odot}$},'' \href{https://dx.doi.org/10.1103/PhysRevLett.125.101102}{{\em
  Phys. Rev. Lett.} {\bfseries 125} no.~10, (2020) 101102},
  \href{https://arxiv.org/abs/2009.01075}{{\ttfamily arXiv:2009.01075
  [gr-qc]}}.

\bibitem{EventHorizonTelescope:2019dse}
{\bfseries Event Horizon Telescope} Collaboration, K.~Akiyama {\em et~al.},
  ``{First M87 Event Horizon Telescope Results. I. The Shadow of the
  Supermassive Black Hole},''
  \href{https://dx.doi.org/10.3847/2041-8213/ab0ec7}{{\em Astrophys. J. Lett.}
  {\bfseries 875} (2019) L1},
  \href{https://arxiv.org/abs/1906.11238}{{\ttfamily arXiv:1906.11238
  [astro-ph.GA]}}.

\bibitem{Christodoulou:1970wf}
D.~Christodoulou, ``{Reversible and irreversible transforations in black hole
  physics},'' \href{https://dx.doi.org/10.1103/PhysRevLett.25.1596}{{\em Phys.
  Rev. Lett.} {\bfseries 25} (1970) 1596--1597}.

\bibitem{Penrose:1969pc}
R.~Penrose, ``{Gravitational collapse: The role of general relativity},''
  \href{https://dx.doi.org/10.1023/A:1016578408204}{{\em Riv. Nuovo Cim.}
  {\bfseries 1} (1969) 252--276}.

\bibitem{Bardeen:1972fi}
J.~M. Bardeen, W.~H. Press, and S.~A. Teukolsky, ``Rotating black holes:
  Locally nonrotating frames, energy extraction, and scalar synchrotron
  radiation,'' {\em Astrophys. J.} {\bfseries 178} (1972) 347.

\bibitem{Wald:1974kya}
R.~M. Wald, ``{Energy Limits on the Penrose Process},''
  \href{https://dx.doi.org/10.1086/152959}{{\em Astrophys. J.} {\bfseries 191}
  (1974) 231}.

\bibitem{Teukolsky:1974yv}
S.~A. Teukolsky and W.~H. Press, ``{Perturbations of a rotating black hole. III
  - Interaction of the hole with gravitational and electromagnet ic
  radiation},'' \href{https://dx.doi.org/10.1086/153180}{{\em Astrophys. J.}
  {\bfseries 193} (1974) 443--461}.

\bibitem{piran1975high}
T.~Piran, J.~Shaham, and J.~Katz, ``High efficiency of the penrose mechanism
  for particle collisions,'' {\em Astrophys. J. Lett.} {\bfseries 196} (1975)
  L107.

\bibitem{Blandford:1977ds}
R.~D. Blandford and R.~L. Znajek, ``{Electromagnetic extractions of energy from
  Kerr black holes},'' \href{https://dx.doi.org/10.1093/mnras/179.3.433}{{\em
  Mon. Not. Roy. Astron. Soc.} {\bfseries 179} (1977) 433--456}.

\bibitem{1990ApJ...363..206T}
M.~{Takahashi}, S.~{Nitta}, Y.~{Tatematsu}, and A.~{Tomimatsu},
  ``{Magnetohydrodynamic Flows in Kerr Geometry: Energy Extraction from Black
  Holes},'' \href{https://dx.doi.org/10.1086/169331}{{\em \apj} {\bfseries 363}
  (Nov., 1990) 206}.

\bibitem{Lee:1999se}
H.~K. Lee, R.~A. M.~J. Wijers, and G.~E. Brown, ``{The Blandford-Znajek process
  as a central engine for a gamma-ray burst},''
  \href{https://dx.doi.org/10.1016/S0370-1573(99)00084-8}{{\em Phys. Rept.}
  {\bfseries 325} (2000) 83--114},
  \href{https://arxiv.org/abs/astro-ph/9906213}{{\ttfamily
  arXiv:astro-ph/9906213}}.

\bibitem{Tchekhovskoy:2008gq}
A.~Tchekhovskoy, J.~C. McKinney, and R.~Narayan, ``{Simulations of
  Ultrarelativistic Magnetodynamic Jets from Gamma-ray Burst Engines},''
  \href{https://dx.doi.org/10.1111/j.1365-2966.2008.13425.x}{{\em Mon. Not.
  Roy. Astron. Soc.} {\bfseries 388} (2008) 551},
  \href{https://arxiv.org/abs/0803.3807}{{\ttfamily arXiv:0803.3807
  [astro-ph]}}.

\bibitem{Komissarov:2009dn}
S.~S. Komissarov and M.~V. Barkov, ``{Activation of the Blandford-Znajek
  mechanism in collapsing stars},''
  \href{https://dx.doi.org/10.1111/j.1365-2966.2009.14831.x}{{\em Mon. Not.
  Roy. Astron. Soc.} {\bfseries 397} (2009) 1153},
  \href{https://arxiv.org/abs/0902.2881}{{\ttfamily arXiv:0902.2881
  [astro-ph.HE]}}.

\bibitem{McKinney:2004ka}
J.~C. McKinney and C.~F. Gammie, ``{A Measurement of the electromagnetic
  luminosity of a Kerr black hole},''
  \href{https://dx.doi.org/10.1086/422244}{{\em Astrophys. J.} {\bfseries 611}
  (2004) 977--995}, \href{https://arxiv.org/abs/astro-ph/0404512}{{\ttfamily
  arXiv:astro-ph/0404512}}.

\bibitem{Hawley:2005xs}
J.~F. Hawley and J.~H. Krolik, ``{Magnetically driven jets in the kerr
  metric},'' \href{https://dx.doi.org/10.1086/500385}{{\em Astrophys. J.}
  {\bfseries 641} (2006) 103--116},
  \href{https://arxiv.org/abs/astro-ph/0512227}{{\ttfamily
  arXiv:astro-ph/0512227}}.

\bibitem{Komissarov:2007rc}
S.~S. Komissarov and J.~C. McKinney, ``{Meissner effect and Blandford-Znajek
  mechanism in conductive black hole magnetospheres},''
  \href{https://dx.doi.org/10.1111/j.1745-3933.2007.00301.x}{{\em Mon. Not.
  Roy. Astron. Soc.} {\bfseries 377} (2007) L49--L53},
  \href{https://arxiv.org/abs/astro-ph/0702269}{{\ttfamily
  arXiv:astro-ph/0702269}}.

\bibitem{Tchekhovskoy:2011zx}
A.~Tchekhovskoy, R.~Narayan, and J.~C. McKinney, ``{Efficient Generation of
  Jets from Magnetically Arrested Accretion on a Rapidly Spinning Black
  Hole},'' \href{https://dx.doi.org/10.1111/j.1745-3933.2011.01147.x}{{\em Mon.
  Not. Roy. Astron. Soc.} {\bfseries 418} (2011) L79--L83},
  \href{https://arxiv.org/abs/1108.0412}{{\ttfamily arXiv:1108.0412
  [astro-ph.HE]}}.

\bibitem{Koide:2008xr}
S.~Koide and K.~Arai, ``Energy extraction from a rotating black hole by
  magnetic reconnection in ergosphere,'' {\em Astrophys. J.} {\bfseries 682}
  (2008) 1124, \href{https://arxiv.org/abs/0805.0044}{{\ttfamily
  arxiv:0805.0044 [astro-ph]}}.

\bibitem{Comisso:2020ykg}
L.~Comisso and F.~A. Asenjo, ``Magnetic reconnection as a mechanism for energy
  extraction from rotating black holes,'' {\em Phys. Rev. D} {\bfseries 103}
  no.~2, (2021) 023014, \href{https://arxiv.org/abs/2012.00879}{{\ttfamily
  arxiv:2012.00879 [astro-ph.HE]}}.

\bibitem{Parfrey:2018dnc}
K.~Parfrey, A.~Philippov, and B.~Cerutti, ``{First-Principles Plasma
  Simulations of Black-Hole Jet Launching},''
  \href{https://dx.doi.org/10.1103/PhysRevLett.122.035101}{{\em Phys. Rev.
  Lett.} {\bfseries 122} no.~3, (2019) 035101},
  \href{https://arxiv.org/abs/1810.03613}{{\ttfamily arXiv:1810.03613
  [astro-ph.HE]}}.

\bibitem{Komissarov:2005wj}
S.~S. Komissarov, ``{Observations of the Blandford-Znajek and the MHD Penrose
  processes in computer simulations of black hole magnetospheres},''
  \href{https://dx.doi.org/10.1111/j.1365-2966.2005.08974.x}{{\em Mon. Not.
  Roy. Astron. Soc.} {\bfseries 359} (2005) 801--808},
  \href{https://arxiv.org/abs/astro-ph/0501599}{{\ttfamily
  arXiv:astro-ph/0501599}}.

\bibitem{East:2018ayf}
W.~E. East and H.~Yang, ``{Magnetosphere of a spinning black hole and the role
  of the current sheet},''
  \href{https://dx.doi.org/10.1103/PhysRevD.98.023008}{{\em Phys. Rev. D}
  {\bfseries 98} no.~2, (2018) 023008},
  \href{https://arxiv.org/abs/1805.05952}{{\ttfamily arXiv:1805.05952
  [astro-ph.HE]}}.

\bibitem{Ripperda:2020bpz}
B.~Ripperda, F.~Bacchini, and A.~Philippov, ``{Magnetic Reconnection and Hot
  Spot Formation in Black Hole Accretion Disks},''
  \href{https://dx.doi.org/10.3847/1538-4357/ababab}{{\em Astrophys. J.}
  {\bfseries 900} no.~2, (2020) 100},
  \href{https://arxiv.org/abs/2003.04330}{{\ttfamily arXiv:2003.04330
  [astro-ph.HE]}}.

\bibitem{comisso2016General}
L.~Comisso, M.~Lingam, Y.-M. Huang, and A.~Bhattacharjee, ``General theory of
  the plasmoid instability,'' {\em Physics of Plasmas} {\bfseries 23} no.~10,
  (2016) 100702, \href{https://arxiv.org/abs/1608.04692}{{\ttfamily
  arxiv:1608.04692 [astro-ph, physics:math-ph, physics:physics]}}.

\bibitem{Uzdensky:2010ts}
D.~A. Uzdensky, N.~F. Loureiro, and A.~A. Schekochihin, ``Fast magnetic
  reconnection in the plasmoid-dominated regime,'' {\em Phys. Rev. Lett.}
  {\bfseries 105} (2010) 235002,
  \href{https://arxiv.org/abs/1008.3330}{{\ttfamily arxiv:1008.3330
  [astro-ph.SR]}}.

\bibitem{Comisso:2017arh}
L.~Comisso, M.~Lingam, Y.-M. Huang, and A.~Bhattacharjee, ``Plasmoid
  instability in forming current sheets,'' {\em Astrophys. J.} {\bfseries 850}
  no.~2, (2017) 142, \href{https://arxiv.org/abs/1707.01862}{{\ttfamily
  arxiv:1707.01862 [astro-ph.HE]}}.

\bibitem{daughton2009Transition}
W.~Daughton, V.~Roytershteyn, B.~J. Albright, H.~Karimabadi, L.~Yin, and K.~J.
  Bowers, ``Transition from collisional to kinetic regimes in large-scale
  reconnection layers,'' {\em Phys. Rev. Lett.} {\bfseries 103} no.~6, (2009)
  065004.

\bibitem{bhattacharjee2009Fast}
A.~Bhattacharjee, Y.-M. Huang, H.~Yang, and B.~Rogers, ``Fast reconnection in
  high-lundquist-number plasmas due to secondary tearing instabilities,'' {\em
  Physics of Plasmas} {\bfseries 16} no.~11, (2009) 112102,
  \href{https://arxiv.org/abs/0906.5599}{{\ttfamily arxiv:0906.5599
  [physics]}}.

\bibitem{Mori:2013yda}
K.~Mori {\em et~al.}, ``{NuSTAR discovery of a 3.76-second transient magnetar
  near Sagittarius A*},''
  \href{https://dx.doi.org/10.1088/2041-8205/770/2/L23}{{\em Astrophys. J.
  Lett.} {\bfseries 770} (2013) L23},
  \href{https://arxiv.org/abs/1305.1945}{{\ttfamily arXiv:1305.1945
  [astro-ph.HE]}}.

\bibitem{Kennea:2013dfa}
J.~A. Kennea {\em et~al.}, ``{Swift Discovery of a New Soft Gamma Repeater, SGR
  J1745-29, near Sagittarius A*},''
  \href{https://dx.doi.org/10.1088/2041-8205/770/2/L24}{{\em Astrophys. J.
  Lett.} {\bfseries 770} (2013) L24},
  \href{https://arxiv.org/abs/1305.2128}{{\ttfamily arXiv:1305.2128
  [astro-ph.HE]}}.

\bibitem{Eatough:2013nva}
R.~P. Eatough {\em et~al.}, ``{A strong magnetic field around the supermassive
  black hole at the centre of the Galaxy},''
  \href{https://dx.doi.org/10.1038/nature12499}{{\em Nature} {\bfseries 501}
  (2013) 391--394}, \href{https://arxiv.org/abs/1308.3147}{{\ttfamily
  arXiv:1308.3147 [astro-ph.GA]}}.

\bibitem{Olausen:2013bpa}
S.~A. Olausen and V.~M. Kaspi, ``{The McGill Magnetar Catalog},''
  \href{https://dx.doi.org/10.1088/0067-0049/212/1/6}{{\em Astrophys. J.
  Suppl.} {\bfseries 212} (2014) 6},
  \href{https://arxiv.org/abs/1309.4167}{{\ttfamily arXiv:1309.4167
  [astro-ph.HE]}}.

\bibitem{EventHorizonTelescope:2021srq}
{\bfseries Event Horizon Telescope} Collaboration, K.~Akiyama {\em et~al.},
  ``{First M87 Event Horizon Telescope Results. VIII. Magnetic Field Structure
  near The Event Horizon},''
  \href{https://dx.doi.org/10.3847/2041-8213/abe4de}{{\em Astrophys. J. Lett.}
  {\bfseries 910} no.~1, (2021) L13},
  \href{https://arxiv.org/abs/2105.01173}{{\ttfamily arXiv:2105.01173
  [astro-ph.HE]}}.

\bibitem{Kanti:1995vq}
P.~Kanti, N.~E. Mavromatos, J.~Rizos, K.~Tamvakis, and E.~Winstanley,
  ``{Dilatonic black holes in higher curvature string gravity},''
  \href{https://dx.doi.org/10.1103/PhysRevD.54.5049}{{\em Phys. Rev. D}
  {\bfseries 54} (1996) 5049--5058},
  \href{https://arxiv.org/abs/hep-th/9511071}{{\ttfamily
  arXiv:hep-th/9511071}}.

\bibitem{Ayzenberg:2014aka}
D.~Ayzenberg and N.~Yunes, ``{Slowly-Rotating Black Holes in
  Einstein-Dilaton-Gauss-Bonnet Gravity: Quadratic Order in Spin Solutions},''
  \href{https://dx.doi.org/10.1103/PhysRevD.90.044066}{{\em Phys. Rev. D}
  {\bfseries 90} (2014) 044066},
  \href{https://arxiv.org/abs/1405.2133}{{\ttfamily arXiv:1405.2133 [gr-qc]}}.
  [Erratum: Phys.Rev.D 91, 069905 (2015)].

\bibitem{Maselli:2015tta}
A.~Maselli, P.~Pani, L.~Gualtieri, and V.~Ferrari, ``{Rotating black holes in
  Einstein-Dilaton-Gauss-Bonnet gravity with finite coupling},''
  \href{https://dx.doi.org/10.1103/PhysRevD.92.083014}{{\em Phys. Rev. D}
  {\bfseries 92} no.~8, (2015) 083014},
  \href{https://arxiv.org/abs/1507.00680}{{\ttfamily arXiv:1507.00680
  [gr-qc]}}.

\bibitem{Kleihaus:2015aje}
B.~Kleihaus, J.~Kunz, S.~Mojica, and E.~Radu, ``{Spinning black holes in
  Einstein\textendash{}Gauss-Bonnet\textendash{}dilaton theory: Nonperturbative
  solutions},'' \href{https://dx.doi.org/10.1103/PhysRevD.93.044047}{{\em Phys.
  Rev. D} {\bfseries 93} no.~4, (2016) 044047},
  \href{https://arxiv.org/abs/1511.05513}{{\ttfamily arXiv:1511.05513
  [gr-qc]}}.

\bibitem{Kokkotas:2017ymc}
K.~D. Kokkotas, R.~A. Konoplya, and A.~Zhidenko, ``{Analytical approximation
  for the Einstein-dilaton-Gauss-Bonnet black hole metric},''
  \href{https://dx.doi.org/10.1103/PhysRevD.96.064004}{{\em Phys. Rev. D}
  {\bfseries 96} no.~6, (2017) 064004},
  \href{https://arxiv.org/abs/1706.07460}{{\ttfamily arXiv:1706.07460
  [gr-qc]}}.

\bibitem{Yunes:2009hc}
N.~Yunes and F.~Pretorius, ``{Dynamical Chern-Simons Modified Gravity. I.
  Spinning Black Holes in the Slow-Rotation Approximation},''
  \href{https://dx.doi.org/10.1103/PhysRevD.79.084043}{{\em Phys. Rev. D}
  {\bfseries 79} (2009) 084043},
  \href{https://arxiv.org/abs/0902.4669}{{\ttfamily arXiv:0902.4669 [gr-qc]}}.

\bibitem{Yagi:2012ya}
K.~Yagi, N.~Yunes, and T.~Tanaka, ``{Slowly Rotating Black Holes in Dynamical
  Chern-Simons Gravity: Deformation Quadratic in the Spin},''
  \href{https://dx.doi.org/10.1103/PhysRevD.86.044037}{{\em Phys. Rev. D}
  {\bfseries 86} (2012) 044037},
  \href{https://arxiv.org/abs/1206.6130}{{\ttfamily arXiv:1206.6130 [gr-qc]}}.
  [Erratum: Phys.Rev.D 89, 049902 (2014)].

\bibitem{McNees:2015srl}
R.~McNees, L.~C. Stein, and N.~Yunes, ``{Extremal black holes in dynamical
  Chern\textendash{}Simons gravity},''
  \href{https://dx.doi.org/10.1088/0264-9381/33/23/235013}{{\em Class. Quant.
  Grav.} {\bfseries 33} no.~23, (2016) 235013},
  \href{https://arxiv.org/abs/1512.05453}{{\ttfamily arXiv:1512.05453
  [gr-qc]}}.

\bibitem{Delsate:2018ome}
T.~Delsate, C.~Herdeiro, and E.~Radu, ``{Non-perturbative spinning black holes
  in dynamical Chern\textendash{}Simons gravity},''
  \href{https://dx.doi.org/10.1016/j.physletb.2018.09.060}{{\em Phys. Lett. B}
  {\bfseries 787} (2018) 8--15},
  \href{https://arxiv.org/abs/1806.06700}{{\ttfamily arXiv:1806.06700
  [gr-qc]}}.

\bibitem{Sen:1992ua}
A.~Sen, ``{Rotating charged black hole solution in heterotic string theory},''
  \href{https://dx.doi.org/10.1103/PhysRevLett.69.1006}{{\em Phys. Rev. Lett.}
  {\bfseries 69} (1992) 1006--1009},
  \href{https://arxiv.org/abs/hep-th/9204046}{{\ttfamily
  arXiv:hep-th/9204046}}.

\bibitem{Vigeland:2011ji}
S.~Vigeland, N.~Yunes, and L.~Stein, ``{Bumpy Black Holes in Alternate Theories
  of Gravity},'' \href{https://dx.doi.org/10.1103/PhysRevD.83.104027}{{\em
  Phys. Rev. D} {\bfseries 83} (2011) 104027},
  \href{https://arxiv.org/abs/1102.3706}{{\ttfamily arXiv:1102.3706 [gr-qc]}}.

\bibitem{Johannsen:2013szh}
T.~Johannsen, ``{Regular Black Hole Metric with Three Constants of Motion},''
  \href{https://dx.doi.org/10.1103/PhysRevD.88.044002}{{\em Phys. Rev. D}
  {\bfseries 88} no.~4, (2013) 044002},
  \href{https://arxiv.org/abs/1501.02809}{{\ttfamily arXiv:1501.02809
  [gr-qc]}}.

\bibitem{Konoplya:2016jvv}
R.~Konoplya, L.~Rezzolla, and A.~Zhidenko, ``General parametrization of
  axisymmetric black holes in metric theories of gravity,'' {\em Phys. Rev. D}
  {\bfseries 93} no.~6, (2016) 064015,
  \href{https://arxiv.org/abs/1602.02378}{{\ttfamily arxiv:1602.02378
  [gr-qc]}}.

\bibitem{Papadopoulos:2018nvd}
G.~O. Papadopoulos and K.~D. Kokkotas, ``{Preserving Kerr symmetries in
  deformed spacetimes},''
  \href{https://dx.doi.org/10.1088/1361-6382/aad7f4}{{\em Class. Quant. Grav.}
  {\bfseries 35} no.~18, (2018) 185014},
  \href{https://arxiv.org/abs/1807.08594}{{\ttfamily arXiv:1807.08594
  [gr-qc]}}.

\bibitem{Carson:2020dez}
Z.~Carson and K.~Yagi, ``{Asymptotically flat, parameterized black hole metric
  preserving Kerr symmetries},''
  \href{https://dx.doi.org/10.1103/PhysRevD.101.084030}{{\em Phys. Rev. D}
  {\bfseries 101} no.~8, (2020) 084030},
  \href{https://arxiv.org/abs/2002.01028}{{\ttfamily arXiv:2002.01028
  [gr-qc]}}.

\bibitem{Wei:2022jbi}
S.-W. Wei, H.-M. Wang, Y.-P. Zhang, and Y.-X. Liu, ``Effects of tidal charge on
  magnetic reconnection and energy extraction from spinning braneworld black
  hole,'' {\em JCAP} {\bfseries 04} no.~04, (2022) 050,
  \href{https://arxiv.org/abs/2201.12729}{{\ttfamily arxiv:2201.12729
  [gr-qc]}}.

\bibitem{Liu:2022qnr}
W.~Liu, ``Energy extraction via magnetic reconnection in the ergosphere of a
  rotating non-kerr black hole,'' {\em Astrophys. J.} {\bfseries 925} no.~2,
  (2022) 149, \href{https://arxiv.org/abs/2204.07338}{{\ttfamily
  arxiv:2204.07338 [astro-ph.HE]}}.

\bibitem{Carleo:2022qlv}
A.~Carleo, G.~Lambiase, and L.~Mastrototaro, ``Energy extraction via magnetic
  reconnection in lorentz breaking kerr{\textendash}sen and kiselev black
  holes,'' {\em Eur. Phys. J. C} {\bfseries 82} no.~9, (2022) 776,
  \href{https://arxiv.org/abs/2206.12988}{{\ttfamily arxiv:2206.12988
  [gr-qc]}}.

\bibitem{Khodadi:2022dff}
M.~Khodadi, ``Magnetic reconnection and energy extraction from a spinning black
  hole with broken lorentz symmetry,'' {\em Phys. Rev. D} {\bfseries 105}
  no.~2, (2022) 023025, \href{https://arxiv.org/abs/2201.02765}{{\ttfamily
  arxiv:2201.02765 [gr-qc]}}.

\bibitem{Li:2023nmy}
Z.~Li, X.-K. Guo, and F.~Yuan, ``Energy extraction from rotating regular black
  hole via comisso-asenjo mechanism,'' {\em Phys. Rev. D} {\bfseries 108}
  no.~4, (2023) 044067, \href{https://arxiv.org/abs/2304.08831}{{\ttfamily
  arxiv:2304.08831 [gr-qc]}}.

\bibitem{Li:2023htz}
Z.~Li and F.~Yuan, ``Energy extraction via comisso-asenjo mechanism from
  rotating hairy black hole,'' {\em Phys. Rev. D} {\bfseries 108} no.~2, (2023)
  024039, \href{https://arxiv.org/abs/2304.12553}{{\ttfamily arxiv:2304.12553
  [gr-qc]}}.

\bibitem{Khodadi:2023juk}
M.~Khodadi, D.~F. Mota, and A.~Sheykhi, ``Harvesting energy driven by
  comisso-asenjo process from kerr-mog black holes,'' {\em JCAP} {\bfseries 10}
  (2023) 034, \href{https://arxiv.org/abs/2307.00478}{{\ttfamily
  arxiv:2307.00478 [astro-ph.HE]}}.

\bibitem{Shaymatov:2023dtt}
S.~Shaymatov, M.~Alloqulov, B.~Ahmedov, and A.~Wang, ``A kerr-newman-mog black
  hole's impact on the magnetic reconnection,'' {\em arXiv:2307.03012 [gr-qc]}
  (2023) , \href{https://arxiv.org/abs/2307.03012}{{\ttfamily arxiv:2307.03012
  [gr-qc]}}.

\bibitem{Wang:2022qmg}
C.-H. Wang, C.-Q. Pang, and S.-W. Wei, ``Extracting energy via magnetic
  reconnection from kerr{\textendash}de sitter black holes,'' {\em Phys. Rev.
  D} {\bfseries 106} no.~12, (2022) 124050,
  \href{https://arxiv.org/abs/2209.08837}{{\ttfamily arxiv:2209.08837
  [gr-qc]}}.

\bibitem{Ni:2016uik}
Y.~Ni, J.~Jiang, and C.~Bambi, ``Testing the kerr metric with the iron line and
  the krz parametrization,'' {\em JCAP} {\bfseries 09} no.~09, (2016) 014,
  \href{https://arxiv.org/abs/1607.04893}{{\ttfamily arxiv:1607.04893
  [gr-qc]}}.

\bibitem{Nampalliwar:2019iti}
S.~Nampalliwar, S.~Xin, S.~Srivastava, A.~B. Abdikamalov, D.~Ayzenberg,
  C.~Bambi, T.~Dauser, J.~A. Garcia, and A.~Tripathi, ``Testing general
  relativity with x-ray reflection spectroscopy: The konoplya-rezzolla-zhidenko
  parametrization,'' {\em Phys. Rev. D} {\bfseries 102} no.~12, (2020) 124071,
  \href{https://arxiv.org/abs/1903.12119}{{\ttfamily arxiv:1903.12119
  [gr-qc]}}.

\bibitem{Li:2023qcu}
S.~Li and W.-B. Han, ``A full waveform model for arbitrarily axis-symmetric
  black hole mergers,'' {\em Phys. Rev. D} {\bfseries 108} no.~8, (2023)
  083032, \href{https://arxiv.org/abs/2307.00797}{{\ttfamily arxiv:2307.00797
  [gr-qc]}}.

\bibitem{Abdikamalov:2021zwv}
A.~B. Abdikamalov, D.~Ayzenberg, C.~Bambi, S.~Nampalliwar, and A.~Tripathi,
  ``Constraining the krz deformation parameters i: Limits from supermassive
  black hole x-ray data,'' {\em Phys. Rev. D} {\bfseries 104} no.~2, (2021)
  024058, \href{https://arxiv.org/abs/2104.04183}{{\ttfamily arxiv:2104.04183
  [astro-ph.HE]}}.

\bibitem{Konoplya:2021qll}
R.~A. Konoplya, J.~Kunz, and A.~Zhidenko, ``{Blandford-Znajek mechanism in the
  general stationary axially-symmetric black-hole spacetime},''
  \href{https://dx.doi.org/10.1088/1475-7516/2021/12/002}{{\em JCAP} {\bfseries
  12} no.~12, (2021) 002}, \href{https://arxiv.org/abs/2102.10649}{{\ttfamily
  arXiv:2102.10649 [gr-qc]}}.

\bibitem{Shaymatov:2023jfa}
S.~Shaymatov, B.~Ahmedov, M.~De~Laurentis, M.~Jamil, Q.~Wu, A.~Wang, and
  M.~Azreg-A\"\i{}nou, ``{On the Parameters of the Spherically Symmetric
  Parameterized Rezzolla\textendash{}Zhidenko Spacetime through Solar System
  Tests, the Orbit of the S2 Star about Sgr A*, and Quasiperiodic
  Oscillations},'' \href{https://dx.doi.org/10.3847/1538-4357/acfcba}{{\em
  Astrophys. J.} {\bfseries 959} no.~1, (2023) 6},
  \href{https://arxiv.org/abs/2307.10804}{{\ttfamily arXiv:2307.10804
  [gr-qc]}}.

\end{thebibliography}\endgroup
\end{document}